# Molecular Dynamics for Low Temperature Plasma-Surface Interaction Studies

David B. Graves [a)] and Pascal Brault

*Groupe de Recherches sur l'Energétique des Milieux Ionisés*

*UMR6606 CNRS*

*Université d'Orléans BP 6744*

*45067 Orléans Cedex 2 France*

[a)] permanent address: *Department of Chemical Engineering*

*University of California, Berkeley, CA 94720 USA*

graves@berkeley.edu

**Abstract**

The mechanisms of physical and chemical interactions of low temperature plasmas with surfaces can be fruitfully explored using molecular dynamics (MD) simulations. MD simulations follow the detailed motion of sets of interacting atoms through integration of atomic equations of motion, using inter-atomic potentials that can account for bond breaking and formation that result when energetic species from the plasma impact surfaces. This article summarizes the current status of the technique for various applications of low temperature plasmas to material processing technologies. The method is reviewed, and commonly used inter-atomic potentials are described. Special attention is paid to the use of MD in understanding various representative applications, including tetrahedral amorphous carbon film deposition from energetic carbon ions; the interactions of radical species with amorphous hydrogenated silicon films; silicon nano-particles in plasmas; and plasma etching.

# 1. Introduction

Ionized gas plasmas usually interact strongly with their bounding surfaces. This is true even for low temperature plasmas; indeed, it is one of the reasons that such plasmas are so widely used for modifying surfaces. Species from the plasma can dramatically alter the near-surface region through various physical and chemical processes. These effects can be carefully controlled to achieve exquisitely sculpted nanoscale features or extraordinary surface properties, even over surfaces with characteristic lengths of meters. Of course, it is also true that plasma-surface interactions can pose considerable problems as well. Controlled nuclear fusion devices employing highly energetic, hot, magnetized plasmas have yet to be developed for sustained (much less commercial) power production in part because plasma-wall interactions contaminate the burning plasma. [Federici, 2003; Matera, 1996; Shimada, 2005; Roth, 2007] Commercial plasma reactors ('tools') used to deposit or etch thin films for semiconductor device manufacturing are often plagued by wall-related contamination. [Ullal et al., 2002; Kim and Aydil, 2003; Lee, Graves and Lieberman, 1996] The design and operation of plasma thrusters, plasma displays, plasma lamps, plasma lasers, and many other applications of plasmas, are often challenged by plasma-wall interactions. [Bouchoule, et al, 2001; Bœuf, 2003]

The variety of surface effects achievable with plasma exposure is remarkable, but the often subtle, and generally synergistic processes that couple plasma conditions with specific surface alterations are notoriously difficult to identify and control. The fact that plasmas can alter surfaces so dramatically with such a wide range of effects is related to the complexity of the plasmas themselves - the range of species and energies that can be created is huge. The 'parameter space' of processing plasmas is truly enormous, and the problem of finding just the right combination of plasma conditions and equipment design is the seemingly unending task of the plasma process engineer in industry or the materials scientist in the laboratory.

Even if the composition and energy of the species impacting the surface were known exactly, the way they combine with the surface to create the observed effects is not generally understood. With a few exceptions, usually involving carefully simplified laboratory experiments, little is known of the atomic scale *mechanisms* relating the impacting species and their resulting effects at surfaces. This is hardly surprising given the obvious fact that the effects of the impacting species occur at surfaces immersed in the complex plasma environment, and that these surfaces are often difficult to probe *in-situ* and in real time. The complexity of the processes limits the use of purely theoretical methods. The obvious way to proceed is to develop and apply numerical simulations to interpret laboratory measurements. A popular, but not exclusive, technique to simulate plasma-surface interactions is molecular dynamics (MD), for reasons that we describe below.

The primary focus of this topical review is the use of molecular dynamics to understand plasma-surface interaction mechanisms at the atomic scale. Most of the literature that we summarize here is related to applications of MD used to simulate plasma-surface processes for materials processing. The literature related to plasma-surface interactions is large and growing, and the present review is not exhaustive. We have chosen to focus mostly on several reasonably well studied applications to give a sense of the current status of the field. The emphasis here is on the mechanistic understanding achieved by coupling molecular dynamics simulations with experimental measurements. We are most interested in documenting what has been learned from the simulations rather than describing all of the details of the simulation techniques. Nevertheless, we summarize the methods and techniques in common use, and these vary somewhat with the investigator and the nature of the problems examined. It is also important to fully understand the limitations of MD in simulating plasma-surface interactions. For the most part, these limitations are due to the invariably imperfect approximations of interatomic potentials and the limitations associated with accessible length and time scales in MD.

**2. Modeling and Simulation of Surface Chemical Processes**

Before proceeding to MD simulations of plasma-surface interactions, we address very briefly the broader issues of surface chemical process simulation and modeling. This is an enormous field and we only touch a few highlights for some perspective. Surface chemical processes most closely related to plasma-surface interactions include chemical vapor deposition, heterogeneous catalysis and electrochemical processes.

Frenklach provides a useful summary of the general methods used to simulate surface reactions associated with chemical vapor deposition (CVD). [Frenklach, 1998] This article focuses primarily on simulation strategies of diamond carbon CVD and of other complex CVD processes such as compound semiconductors. He discusses thermodynamic models; global kinetic modeling; detailed kinetic modeling and kinetic Monte Carlo; and finally, MD models. Frenklach points out that thermodynamic models suffer from the generally invalid assumption of equilibrium for all steps, although some fast intermediate reactions may approach equilibrium. General kinetic models conserve mass but lack detailed predictive capability for non-trivial multi-step and multi-species reaction mechanisms. Detailed kinetic models, of the sort used commonly to model the dozens to hundreds of species and hundreds to thousands of elementary reactions occurring in flames and atmospheric chemistry, require information on reaction mechanisms and cannot easily deal with steric constraints and non-local effects common in surface chemistry. Finally, MD requires proper potentials and is limited to relatively short length- and time-scales, typically less than a nanosecond and perhaps a few tens of nanometers, depending strongly on the number of atoms simulated, the type of potential used and the trajectory times desired. Frenklach points out that much progress has been made and will continue by cleverly combining and exploiting the strengths of these methods. Cavallotti et al. make similar points about the challenges of addressing the large range of scales in CVD reactors, and include a discussion including reactor-scale phenomena. [Cavallotti et al., 2005]

Halley et al. address the challenges of modeling electrochemical interfaces. [Halley et al., 2000] They point out that the processes that are important in electrochemical-surface interactions span about 10 orders of magnitude, similar to estimates for plasma-surface interaction studies. Another similarity with low temperature plasma-surface interactions is the common formation of thin 'passivation' layers on surfaces during processing/exposure. These authors stress the importance of finding ways to bridge the gaps in time- and length-scale between methods starting from *ab-initio* MD force field calculations through continuum descriptions.

In heterogeneous catalysis, strenuous efforts have been made over the last decade to develop reliable models of surface processes. Andzelm et al. described many of the varied issues facing molecular simulation of catalysis nearly a decade ago. [Anzelm et al., 1999] A more recent study by Reuter, Frinkel and Scheffler shows how combining what these authors term *ab-initio* statistical mechanics can lead to atomically detailed descriptions of heterogeneous catalysis (in this case CO oxidation on crystalline $RuO_2$ (110)) for experimentally-relevant timescales of seconds. [Reuter et al., 2004] This method combines density functional theory (DFT) calculation of selected processes coupled with transition state theory and *'ab-initio'* kinetic Monte Carlo. This is a good example of what can be done rigorously in heterogeneous catalysis simulation, but it also shows that the method works best for relatively simple systems with well-defined elementary processes. For this example, only 26 elementary processes were identified and the adsorption energies and diffusion and reaction energy barriers were computed using DFT. The limitations of kinetic Monte Carlo involving difficulties with steric hindrance and ignorance of elementary reaction mechanisms, noted by Frenklach (above) for most CVD processes, were absent in this model catalysis example.

**3. Molecular Dynamics Simulations**

**3.1 Introduction to MD for plasma-surface interaction studies**

What are MD simulations and how do they compare with other 'particle simulation' methods? MD refers to simulations of the motion of collections of atoms, molecules or some assemblage of discrete particles. The simulated particles (generically, 'molecules') interact with each other through an inter-atomic potential, the negative gradient of which corresponds to the force exerted between molecules, leading to their relative motion. A central question in MD, of course, is the choice of the inter-atomic potential, and we address this issue in greater detail later in the article. Molecules can be simulated in the various 'ensembles' familiar to the student of statistical mechanics, such as the microcanonical ensemble (N,V, E), implying a constant number of molecules (N), constant volume (V) and constant total energy (E). The term 'MD' generally implies simulations of interactions among molecules in condensed phases. MD is essentially deterministic in the sense that with a given set of initial positions and velocities, the molecular motion will be identical for two simulations. In contrast, so-called 'Monte Carlo' methods (recognizing that this term can mean different things) are based on statistical simulations of configurations. A proposed move (often on a pre-defined grid) is tested to see what the trial move will do to the total system energy. Moves that lower total energy are always accepted and moves that result in higher energy are accepted or rejected based on Boltzmann statistics.

Techniques designed to simulate molecules interacting either through Lorentz forces (electric and magnetic fields interacting with a point charge) and/or through brief neutral-neutral 'collisions,' generally employ spatial meshes or grids. Two well-known techniques are 'particle-in-cell' (PIC) and 'Direct Simulation Monte Carlo' (DSMC) methods. [Birdsall and Langdon, 1985; Hockney and Eastwood, 1981; Bird, 1994] PIC techniques are often applied to simulating interactions between charged particles in plasmas. In this class of method, electric and magnetic field forces are interpolated onto numerical grids or meshes based on local sources of net charge and current. Particles respond to these mesh-localized forces, rather than to other particles directly, and move through the mesh. The assumption (corresponding to an 'ideal plasma') is that the localized, fluctuating Coulomb forces from adjacent particles can be neglected in favor of the collective force from all of the charges within the local Debye sphere. DSMC is also a particle-mesh method, but is generally used to treat rarefied flows of neutral gases. In DSMC, neutral molecules are modeled as freely convecting point masses, with statistical treatment of particle-particle scattering collisions within mesh regions. Simulations of partially ionized plasmas combine the methods: 'PIC-MCC,' or particle-in-cell with Monte Carlo collisions. Both PIC and DSMC approximate the behavior of extremely large numbers of particles with a much smaller set of 'super-particles,' the average behavior of which approaches the collective behavior of the real system. Note that 'scattering' implies, as is the case in the gas phase, that particles interact with each other only over a very short period of time compared to their total trajectory. This contrasts with the behavior of molecules in condensed phases, of course, for which the molecules are always interacting with other molecules in their vicinity. MD is designed to simulate these conditions.

The complexity of many body systems interacting with each other requires machine computation, and this type of simulation began when computers became sufficiently powerful that non-trivial results could be obtained in a reasonable time. The first paper to use this idea is generally acknowledged to be the paper by Alder et al. (1957), in which these authors attempted to simulate the behavior of a collection of particles modeled as 'hard spheres.' Studies of thermodynamic and transport properties in condensed phases, especially liquids, remain a staple of MD applications. [Allen and Tildesley, 1987; Haile, 1992; Rapaport, 1995] The idea of using MD to understand condensed matter thermodynamic and transport behavior derives from the statistical physics notion that macroscopic behavior of collections of molecules is controlled by many body particle-particle interactions. The most straightforward way to simulate this behavior is to simply follow the trajectories of a collection of molecules as they collide with each other. If the particle-particle interactions are sufficiently accurately modeled by the chosen potentials, then the behavior of the simulated collection of molecules will match experiments, assuming the simulation includes enough particles interacting over a long enough time. Larger spatial scale particle-particle interactions (e.g. screened electrostatic interactions) and/or interactions that develop over longer time scales (e.g. macromolecular polymer/protein dynamics) challenge the method. Another issue in liquids is the proper treatment of small solvent molecules, especially water, interacting with various solutes.

Plasma-surface interactions are often dominated by the effects of single, energetic species, usually an ion, impacting a surface. The 'surface' consists of an interconnected set of atoms, and the impacting species disrupts these atoms in various ways. The number of surface atoms affected by a single impact depends on the energy of the impacting species as well as the nature of the substrate-substrate and substrate-ion interactions. It is commonly the case that the majority, and perhaps the great majority, of the effects of the energetic impacting species are completed in several picoseconds ($10^{-12}$ s). This fact makes MD an obvious choice for study of ion-surface or 'radiation damage' studies of surfaces since such relatively short time interactions can be conveniently treated with MD [Anderson, 1987]. In order to capture the effects of many impacts, multiple trajectories are simulated. If the surface changes due to the impacts, then multiple trajectories, typically with different, randomly chosen impacts, are simulated. The effects of species impacting with a distribution of energies and angle of impact can be treated with the obvious extensions. As Urbassek [2007] points out in a recent review of MD used for sputtering simulation, the first use of MD to simulate radiation damage in metals appeared only 3 years after the 1957 paper by Alder et al. [Gibson et al., 1960]

Sputtering is the process of eroding a surface by impacting energetic species, leading to the ejection of surface atoms. It was observed in the earliest studies of gas discharges [e.g. Grove 1852] that discharge-induced electrode materials erosion would lead to film deposition on the inside of the vacuum chambers used to contain the discharge. This phenomenon became the basis of an enormously useful and widely used thin solid film deposition technique, and has been reviewed recently in an excellent volume [Behrisch and Eckstein, 2007]. Sputtering is a subset of plasma processing of materials [Jacob and Roth, 2007]. The most common way to simulate sputtering is through a special kind of 'Monte Carlo' simulation that typically assumes a series of binary pair-wise interactions (scattering or displacement) between the impacting ion and the substrate atoms, and between dislodged and lattice atoms in the surface. [Eckstein and Urbassek, 2007] This method and its variants have been quite successfully used to predict sputtering yields, depth of penetration, surface damage, and so on. MD (sometimes referred to as 'classical trajectory simulations') has also been widely used for sputtering simulations, as noted above. [Urbassek, 2007]  Smith, Harrison and Garrison [1989] point out that MD simulations of sputtering are particularly useful for understanding how large, intact molecules can be ejected from surfaces subject to ion bombardment, and for rationalizing energy and angular distributions of sputtered particles.

**3.2 Inter-atomic potentials used in MD**

All MD calculations require good interaction potentials. In the last number of years, there has been much discussion and development of so-called *ab-initio* MD (AIMD), in which some form of electronic structure calculation is performed at each time step in the simulation, re-computing the inter-atomic potentials and therefore the inter-atomic forces 'on the fly.' [e.g. Iftimie et al., 2005; Vach and Brulin, 2005] This offers the obvious advantage, at least in principle, of greater accuracy, but it comes at a considerable computational cost.

For many of the plasma-surface studies we discuss here, although not for all, the computational burden of AIMD is too great to allow realistic cases to be simulated. One major reason for this is that realistic plasma-surface simulations demand typically many different species, with different energies and angle of incidence, etc. It is often the case that the plasma changes the surface dramatically, and simulating this is expensive. In any event, usually many thousands of trajectories must be simulated, and AIMD is impractical for most investigators. The more common strategy, therefore, is to use empirical potentials whose parameters are fit by some combination of experimental data and separate *ab-initio* calculations. Garrison and Srivastava reviewed the state-of-the art in inter-atomic potentials useful for MD simulation of surface chemical reactions. [1995]

If fine details of specific chemical interactions are needed, separate cluster calculations using *ab-initio* methods can be used to investigate some special configurations (such as a transition state between reactants and products) that result from the MD trajectory simulations. A good illustration of this approach is the work of Maroudas and Aydil and their co-workers in their studies of amorphous hydrogenated silicon, described in greater detail below. [e.g. Sriraman et al., 2002] In this section, we review some of the basics of classical inter-atomic potentials and MD simulations.

The simplest inter-atomic potentials are of pair form: the net effect on any given molecule can be described as the sum over all pair-wise interactions. Pair potentials are suitable for monatomic systems with dense structure, e.g. cubic face centered (CFC). Nevertheless, it is often desirable, indeed necessary, to go beyond the pair approximation. This is clearly true for molecules with covalent bonds due to the intrinsic directionality of covalent bonding. Pair potentials cannot represent bonds with directionality.

In the following, we present a non-exhaustive summary of various analytical inter-atomic empirical potentials for pair, 3- body and N-body interactions. Any chosen potential must be capable of predicting, with an appropriate level of accuracy, all fundamental material properties, including elastic constants, lattice parameters, bond strengths and direction, and surface and bulk energies, among others. [Halicioglu et al, 1975, Girifalco et al, 1959]

### 3.2.1 Pair potentials

We include pair potentials in this discussion because they can often be useful for phenomenological assessments, and can sometimes be used to model CFC materials. Among the more popular pair potentials is the Lennard-Jones form, often used for liquids and polymers,

$$V_{ij}(r_{ij}) = 4\varepsilon\left[\left(\frac{\sigma}{r_{ij}}\right)^{12} - \left(\frac{\sigma}{r_{ij}}\right)^{6}\right] \quad (1)$$

Another popular pair potential is the Morse potential,

$$V_{ij}(r_{ij}) = D_0 \left[ \exp\left[-2\alpha(r_{ij}-r_0)^2\right] - 2\exp\left[-\alpha(r_{ij}-r_0)^2\right] \right] \tag{2}$$

These forms are especially useful when spectroscopic data are available to fit the potential parameters. This includes vibrational energy, force constant and equilibrium distance for fitting the binding energy $D_0$, the potential stiffness α and equilibrium distance $r_0$.

The Buckingham potential combines the Morse and Lennard-Jones,

$$V_{ij}(r_{ij}) = A\exp\left(-\frac{r_{ij}}{r_B}\right) - \frac{C_6}{r_{ij}^6} \tag{3}$$

Typical parameters for common elements are summarized in table 1.

If pair potential parameters for compound materials are not directly available, mixing rules can be used to make approximations. As example, the Lorenz-Berthelot mixing rule is suitable for Lennard-Jones potentials of species A and B as follows: $\varepsilon_{AB}$ = $(\varepsilon_A\varepsilon_B)^{1/2}$ and $\sigma_{AB}$ = $(\sigma_A+\sigma_B)/2$ [Brault et al, 2002].

In the case of high energy particles or ions in plasma or laser interactions with surfaces, repulsive pair potentials are effective for describing short range interactions. Molière and Ziegler-Biersack-Littmark (ZBL) potentials are commonly used. These potentials follow a simple universal form for the screened Coulomb potential:

Molière potential:

$$V_M(r_{ij}) = \frac{Z_1 Z_2 e^2}{4\pi\varepsilon_0 r_{ij}} \sum_{i=1}^{3} c_i \exp\left(-d_i \frac{r_{ij}}{a_F}\right) \tag{4}$$

ZBL potential:

$$V_{ZBL}(r_{ij}) = \frac{Z_1 Z_2 e^2}{4\pi\varepsilon_0 r_{ij}} \sum_{i=1}^{4} c_i \exp\left(-d_i \frac{r_{ij}}{a_U}\right) \quad (5)$$

With

$$a_F = \frac{0.83\left(\frac{9\pi^2}{128}\right)^{1/3} a_B}{\left(Z_1^{1/2} + Z_2^{1/2}\right)^{2/3}} \quad \text{and} \quad a_U = \frac{0.8853 a_B}{\left(Z_1^{0.23} + Z_2^{0.23}\right)} \quad \text{with} \quad a_B = 0.529177 \, \text{Å}$$

The other parameters are listed in table 2

Two others pair potentials are popular, especially for describing SiO$_2$. These are of Born-Meyer type and are known as van Beest-Kramer- van Santen (BKS) [van Beest et al, 2000] and Garofalini – Feuston . [Levine et al, 1986]. They take into account long range ionic interactions. They can be written as:

$$V_{ij}(r_{ij}) = \frac{q_i q_j e^2}{4\pi\varepsilon_0 r_{ij}} + A_{ij} \exp(-b_{ij} r_{ij}) - \frac{C_{ij}}{r_{ij}^6} \quad (6)$$

for BKS , and as

$$V_{ij}(r_{ij}) = B_{ij} \exp\left(-\frac{r_{ij}}{r_G}\right) + \frac{q_i q_j e^2}{4\pi\varepsilon_0 r_{ij}} \, erfc\left(\frac{r_{ij}}{\beta_{ij}}\right) \quad (7)$$

for Garofalini.

The BKS potential includes Coulomb and Born-Meyer forms for the repulsive part and an attractive van der Waals interaction. The Garofalini form includes Born-Mayer and screened Coulomb repulsions. Table 3 summarizes the fit parameters for $SiO_2$.

Pair potentials suffer from some important drawbacks. Most importantly, they do not take into account the directionality of covalent bonds, surface energy, or vacancy energy, among others. This results, in part, from the relative simplicity of the potentials. This is especially true for the two parameter Lennard-Jones potential. Nevertheless, they can provide reasonable approximations to bulk energy and equilibrium atomic separation. In some cases, it is possible to obtain predictions of pair correlation functions, structure factors, density, morphology of deposited layer/etched layers and roughness that agree well with measured values. On this basis, qualitative results can sometimes be extracted from MD calculations using pair potentials.

**3.2.2 Three-body potentials**

The use of three body potentials is the simplest way to introduce many-body effects in MD simulations. Three body potentials are often used for treating covalent bonds. One early attempt was formulated by Stillinger and Weber for describing silicon. [Stillinger and Weber, 1985] Then further refinements included SiF bonds [Stillinger and Weber, 1989; Weber and Stillinger, 1990], SiCl [Feil et al., 1993] and also SiH. Further improvements have been made by the Vashishta group for treating materials such as $SiO_2$ [Vashista et al., 1990; 1997], SiC [Shimojo et al, 2000], $Si_3N_4$ [Vashishta et al, 1990; Omeltchenko, 1998], and $SiSe_2$ [Vashishta et al, 1997].

Formally, any function *V* describing the interactions between N particles can be cast in a sum of terms with 1, 2, 3, … n bodies, according to :

$$V(1,...,N) = \sum_{i} V_1(i) + \sum_{i,j>i} V_2(i,j) + \sum_{i,j,k>j>i} V_3(i,j,k) + \cdots + \sum V_n(1,\cdots,n) \quad (8)$$

The $V_1$ term corresponds to interaction with a wall or external potentials. It is often omitted and thus the potentials are written as a combination of pair and 3-body potentials $V_2$ and $V_3$. The potentials are expressed in reduced form after introducing energy and length scales ε and σ, respectively:

$$V_2(r_{ij}) = \varepsilon f_2\left(\frac{r_{ij}}{\sigma}\right), \quad V_3(\vec{r}_i,\vec{r}_j,\vec{r}_k) = \varepsilon f_3\left(\frac{\vec{r}_i}{\sigma},\frac{\vec{r}_j}{\sigma},\frac{\vec{r}_k}{\sigma}\right) \quad (9)$$

The pair contribution $f_2$, reads:

$$f_2(r) = \begin{cases} A(Br^{-p} - r^{-q})\exp\left(\dfrac{1}{r-a}\right) & r < a \\ 0 & r \geq a \end{cases} \quad (10)$$

This form is truncated at r = a without discontinuity, which is of practical value in MD simulations. The 3-body term can be written:

$$f_3(\vec{r}_i,\vec{r}_j,\vec{r}_k) = h(r_{ij},r_{ik},\theta_{jik}) + h(r_{ji},r_{jk},\theta_{ijk}) + h(r_{ki},r_{kj},\theta_{ikj}) \quad (11)$$

With $\theta_{jik}$ being the angle between $\vec{r}_j$ and $\vec{r}_k$ at the vertex i. And if $r_{ij}, r_{ik} < a$, then

$$h(r_{ij},r_{ik},\theta_{jik}) = \lambda \exp\left[\frac{\gamma}{r_{ij}-a} + \frac{\gamma}{r_{ik}-a}\right] \cdot \left(\cos\theta_{jik} + \frac{1}{3}\right)^2, \text{ else h=0 if } r \geq a, \text{ and } \theta_{jik} = \frac{\vec{r}_{ij} \cdot \vec{r}_{ik}}{r_{ij} r_{ik}}.$$

As an example, the parameters for silicon are:

A =7.049556277    B =0.6022245584

p =4 q =0 a =1.80 Å

λ = 21.0 γ =1.20

ε=2.1692 eV σ =2.0951 Å

A recent improvement for the pair function includes the ionic effects. This leads to the Vashishta potentials [Vashishta et al,1990; 1997; Shimojo et al, 2000; Omeltchenko et al, 1998].

$$V_2 = A_{ij}\left(\frac{\sigma_i + \sigma_j}{r_{ij}}\right)^{\eta_{ij}} + \frac{Z_i Z_j}{r_{ij}} e^{-\frac{r_{ij}}{r_{1s}}} - \frac{\frac{1}{2}(\alpha_i Z_j^2 + \alpha_j Z_i^2)}{r_{ij}^4} e^{-\frac{r_{ij}}{r_{4s}}}, \quad r_{ij} < r_c$$

$$V_3 = \sum_{i<j<k} V_{jik}^{(3)}$$

$$V_{jik}^{(3)} = B_{jik} f(r_{ij}, r_{ik})(\cos\theta_{jik} - \cos\bar{\theta}_{jik}) \tag{12}$$

$$f(r_{ij}, r_{ik}) = \exp\left[\frac{l}{r_{ij} - r_{c3}} + \frac{l}{r_{ik} - r_{c3}}\right], \quad r_{ij}, r_{ik} < r_{c3}$$

$$\cos\theta_{jik} = \frac{\vec{r}_{ij} \cdot \vec{r}_{ik}}{r_{ij} r_{ik}}$$

The first term of $V_2$ describes steric repulsion, while the second term is the screened Coulomb repulsion due to charge exchange. The third term is the charge – dipole interaction due to the high polarizability of negative ions [Vashishta et al, 1990; Shimojo et al, 2000]. $r_c$ and $r_{c3}$ is a cutoff radius. For $Si_3N_4$, $SiSe_2$ and $SiO_2$ only triplets AXA or XAX are allowed, but not XXA, AAX, etc. In that case the 3 – body potentials reduce to a sum of pair potentials, because the scalar product $\vec{r}_{ij} \cdot \vec{r}_{ik}$ simply reduces to $x_{ij} \cdot x_{ik} + y_{ij} \cdot y_{ik} + z_{ij} \cdot z_{ik}$. In table 5, the parameters are listed for $SiO_2$ and $Si_3N_4$.

**3.2.3 N-body potentials**

N-body potentials have been built for better taking into account covalent bonds and more generally tetrahedral bonds such as Si-Si, SiC, $SiH_x$, $SiC_xF_y$, C, $CH_x$ and also BN. The main feature (as for the Vashishta potentials) is to introduce a mean angle which will result in an angle distribution peaked at this angle with a more or less broad width. This simple treatment of a geometrical parameter allows to correctly describe the short range order in covalent materials and the bulk energies of various poly-types. One of the most famous and most widely used is the Tersoff potential [Tersoff, 1988a, 1988b, 1989], for which numerous improvements exist. Among these is the well-known Brenner form [Brenner, 1990]. A convenient form of this potential can be written as [Marcos et al, 1999]:

$$V_{ij}(r_{ij}) = f_c(r_{ij})\{V_R(r_{ij}) - b_{ij}V_A(r_{ij})\}, \quad V_i = \frac{1}{2}\sum_{i \neq j} V_{ij} \tag{13}$$

$V_A(r)$ and $V_R(r)$ being the repulsive and attractive part of a Morse potential, respectively:

$$\begin{aligned} V_A(r) &= \frac{D_0}{S-1}\exp\left[-\beta\sqrt{2S}(r - R_0)\right] \\ V_R(r) &= \frac{D_0 S}{S-1}\exp\left[-\beta\sqrt{\frac{2}{S}}(r - R_0)\right] \end{aligned} \tag{14}$$

$f_c$ is a decreasing cutoff function centered at $R$ and with half-width $D$:

$$f_c(r) = \begin{cases} 1 & (r < R-D) \\ \frac{1}{2} - \frac{1}{2}\sin\left[\frac{\pi}{2}\frac{r-R}{D}\right] & R-D < r < R+D \\ 0 & r \geq R+D \end{cases}, \text{ or } f_c(r) = \frac{1}{1+\left(\frac{r-R}{D}\right)} \tag{15}$$

Finally $b_{ij}$ terms include the dependence against $\theta_{ijk}$, the angle between $i$-$j$ and $i$-$k$ bonds:

$$\begin{aligned}
b_{ij} &= (1+\gamma^n \chi_{ij}^n) \\
\chi_{ij} &= \sum_{k(\neq i,j)}^{N} f_c(r_{ik})g(\theta_{ijk})\exp[\lambda^3(r_{ij}-r_{ik})^3] \\
g(\theta_{ijk}) &= 1+\frac{c^2}{d^2}-\frac{c^2}{d^2+(h-\cos\theta_{ijk})^2}
\end{aligned} \qquad (16)$$

with N being the number of interacting neighbors. Parameters for C, Si and BN are listed in table 4. For compounds systems like SiC and SiGe, mixing rules apply [Tersoff, 1988b]. In particular, if we write the potential in the form

$V(r_{ij}) = f_c(r_{ij})[A_{ij}\exp(-\lambda_{ij}r_{ij}) - B_{ij}\exp(-\mu_{ij}r_{ij})]$, then the mixing rules reads : $\lambda_{ij}=\lambda_i+\lambda_j$, $\mu_{ij}=\mu_i+\mu_j$ and $W_{ij}=(W_iW_j)^{1/2}$ with W standing for A, B, R or S [Tersoff, 1988b].

The second moment approximation of the tight binding theory (TB-SMA) [Rosato et al, 1989, Cleri et al 1993] is well suited for treating bulk and surface properties of transition metal elements since it leads to an analytical form of the potentials. For higher order moment, the potentials are no longer analytical, but it is possible to treat covalent bonding (Si and C), electronic structure, and spd hybridation [Wang et al, 1990]. The TB-SMA approximation also allows deducing the properties of alloys. For example, many applications deal with alloy segregation and organization [Barreteau et al, 2000; Treglia et al, 1999; Goyhenex et al, 2001]. The TB-SMA analytical form can be written as:

$$V_i = \sum_{i\neq j} A\exp\left[-p\left(\frac{r_{ij}}{r_0}-1\right)\right] - \left\{\sum_{i\neq j}\xi^2 \exp\left[-2q\left(\frac{r_{ij}}{r_0}-1\right)\right]\right\}^{\frac{1}{2}} \qquad (17)$$

The first term is a repulsive Born-Meyer potential. The second term is an attractive band energy. In this expression of the potential, $\xi$ is an effective hopping integral; p and q describe radial dependence of the effective hoping integral and repulsive pair bonding; and $r_0$ is the first neighbor distance. As a first approximation, the following relations can be used for finding A and ξ:

$$\xi = \frac{p}{p-q}\frac{E_c}{\sqrt{Z}} \quad and \quad A = q\frac{E_c}{Z} \tag{18}$$

with $E_c$ being the bulk energy and $Z$ the coordination number. Parameters for some common metals are summarized in tables 6 and 7.

A popular technique known as the embedded atom method (EAM) was introduced by Daw and Baskes [Daw and Baskes, 1983, 1984; Foiles et al., 1986; Daw et al., 1993]. It is based on concepts from density functional theory, which stipulates in general that the energy of a solid is a unique function of electron density. In this case, the relevant electron density is assumed to be the local density at each atomic site. The bulk electron density in this limit is approximated by the sum of electron density associated with each atomic site. Thus, the total energy $E_{tot}$ of an atomic system can be expressed as:

$$E_{tot} = \sum_i V_i, \quad V_i = \frac{1}{2}\sum_{j\neq i}\phi_{ij}(r_{ij}) + F(\rho_i), \quad with \quad \rho_i = \sum_{j\neq i} f_i(r_{ij}) \tag{19}$$

In (19), $V_i$ is the internal atom energy, $\rho_i$ is the electron density associated with atom $i$ due to the presence of other atoms in the system. $F(\rho_i)$ is the energy required to 'embed' the atom $i$ in the electron density $\rho_i$. $\phi_{ij}(r_{ij})$ is a pair potential that can be chosen to be as simple as possible: generalized Lennard-Jones, Morse, Born-Meyer, $\frac{a}{r^n}$, etc. When $\phi_{ij}(r_{ij})$ is of Born-Meyer form and $F(\rho_i) = -\left\{\sum_{i\neq j}\xi^2 \exp\left[-2q\left(\frac{r_{ij}}{r_0}-1\right)\right]\right\}^{\frac{1}{2}}$, the TB-SMA approximation is recovered. If $F(\rho_i) \propto \rho_i^{1/2}$ and $\rho_i \propto \sum_j \left(\frac{a_{ij}}{r_{ij}}\right)^m$, one find the Finnis-Sinclair [Finnis and Sinclair, 1984] and Sutton-Chen [Sutton and Chen, 1990] potentials, respectively.

**3.3 MD simulation procedures**

Molecular dynamics simulations involve numerically solving equations of motion of a set of N interacting atoms or molecules [Allen and Tildsley, 1987; Frenkel and Smit, 1996; Rapaport, 1998]. For Hamiltonian systems, this reduces to solving Newton's second law. If dissipation occurs, through friction terms for example, Langevin-like equations have to be solved. For simplicity, we consider the cases for which Newton's equations of motion are valid. They can be written:

$$\frac{\partial^2 \vec{r}_i(t)}{\partial t^2} = \frac{1}{m_i}\vec{f}_i, \quad \text{with the force } \vec{f}_i = -\frac{\partial}{\partial \vec{r}}V(\vec{r}_1(t), \vec{r}_2(t), \cdots, \vec{r}_N(t)) \qquad (20)$$

The only information necessary to solve this set of N equations of motion is the potential energy $V(\vec{r}_1, \vec{r}_2, \cdots, \vec{r}_N)$. Statistical information and materials properties can be deduced by averaging over all trajectories $\vec{r}_i(t)$.

Many methods are available for integrating these sets of equations. The most popular are finite difference techniques with or without variable time step and/or space step. In Hamiltonian systems, the criterion of quality is generally energy conservation. Conservation of this quantity is ensured by a proper value of time step $\Delta t$. Typically, it should be of the order of $10^{-15}$ s for kinetic energy below 1eV. For higher kinetic energies, the following rule [Beardmore et al, 1998] can be adopted:

$$dt = \frac{C}{\sqrt{\max\limits_{1 \leq i \leq N}\left(\frac{2[E_{kin} + \max(0, V_i)]}{m_i}\right)}}$$ with $C$ being a constant of order of 0.1 Å.

The most time consuming part of the calculation is the force evaluation at each time step. For example, for a pair potential the CPU times generally evolves as $O(N^2)$ and for N-body interaction it becomes $O(N^N)$. This can be improved by properly selecting the relevant interacting particles. For pair interaction the cost can be reduced to a linear dependence with the particle number N by using the linked cell list method, described below [Allen and Tildsley, 1987; Frenkel and Smit, 1996]. For long range Coulomb interactions, some algorithms have been developed by using "divide and conquer" methods, like the Fast Multipole Method (FMM) [Greengard et al, 1987]. Multi-timestep methods can also reduce the computer time [Streett et al, 1978; Tuckermann et al, 1992]. In this case the potential is divided into parts: the first part (for short range interactions) is evaluated at each time step, while the long range part is calculated after a few time steps.

### 3.3.1 Link cell list

The linked cell list method consists in dividing the simulation box into cells. The number of atoms in each cell is saved in a linked list [Allen and Tildsley, 1987; Frenkel and Smit, 1996]. The size $d$ of each cell in the real space is chosen to be $d = r_c + \delta$ where $r_c$ is the potential cutoff radius and $\delta$ is a 'skin depth.' To calculate the force acting on particle i, we only need to sum the contributions of particles in the cell and in the neighboring cells. In two dimensions there are 9 relevant cells while in 3 dimensions there are 27 relevant cells. The calculation now becomes proportional to the mean number MN of atoms relevant for force calculation. If $\rho$ is the mean atomic density, M = $\rho(3d)^3 = 27 \rho (r_c + \delta)^3$ (in 3D). Thus the algorithm computation time scales as O(N) for pair interactions. Exploitation of Newton's third law $f_{ij} = -f_{ji}$, further lowers CPU time by a factor 2.

The skin depth $\delta$ allows updating the list after some time steps corresponding to a length $\delta/2$. This can be further improved by using a cell size $d = (r_c + \delta)/k$ where $k$ is the number of cells per cutoff radius. Thus the number of relevant neighbors becomes : M = $\rho((2k+1)d)^3 = \rho (2+1/k)^3 (r_c + \delta)^3$ (in 3D). If k=2, then the time cost is reduced by a factor ~ 1.7. The condition on cell size is only that the cell must contain enough particles. Otherwise, too much time is spent looping over empty cells. The linked list can be used to determine the neighbor list of each atom $i$, which needs to be updated only when the particles have moved more than a distance $\delta/2$. Thus M = $4/3\pi \rho(r_c + \delta)^3 \sim 4.2 \rho(r_c + \delta)^3$. Using again Newton's third law divides the CPU time by 2. The main problem of the linked cell list scheme is the storage of information on the positions and momenta of MN particles, which can become unmanageably large for long range potentials.

### 3.2 Solving equations of motion

A large set of methods is available for integrating the equations of motion (20). Because the force calculation is time consuming, it is preferable to choose a method allowing a relatively large time-step. An efficient algorithm should be able to provide a good approximation of the solution on short time scales and ensuring conservation of total energy over large timescales. The 'velocity Verlet' algorithm is known to fulfill these stability conditions at large time scales. [Swope et al, 1982]. This algorithm works as follows: assuming that positions $\vec{r}_i(t)$, velocities $\vec{v}_i(t)$ and accelerations $\vec{a}_i(t) = -\frac{\vec{f}_i(t)}{m_i}$ are known at time $t$, the velocities $\vec{v}_i(t + \frac{dt}{2})$ at time t+dt/2 can be evaluated by :

$\vec{v}_i(t + \frac{dt}{2}) = \vec{v}_i(t) + \vec{a}_i(t)\frac{dt}{2}$. Next, the new positions at time $t+dt$ can be estimated as

$\vec{r}_i(t + dt) = \vec{r}_i(t) + \vec{v}_i(t + \frac{dt}{2})$. Accelerations at t+dt $\vec{a}_i(t + dt) = \vec{a}_i[\vec{r}_i(t + dt)]$ can be calculated and the velocities at time t+dt can be determined:

$\vec{v}_i(t + dt) = \vec{v}_i(t + \frac{dt}{2}) + \vec{a}_i(t + dt)\frac{dt}{2}$. This is repeated at each time step.

### 3.3.3 Some peculiarities of MD simulations of interactions with surfaces

Simulations of thin film deposition, etching or material treatment using MD require some additional comments. Bonding or interactions with the surface require correctly treating the energy exchanges between incoming atoms (or ions) and the surface atoms. In the case of ions impacting surfaces, it is generally assumed that ions recombine via Auger neutralization before impacting the surface, so potentials for neutrals can be used. In addition, the cells used to approximate surfaces generally use periodic lateral boundaries and fixed layers of atoms at the bottom of the cell. Energy deposited at the surface by energetic species impact must eventually be removed to maintain the cell at the desired temperature. An 'exact' treatment of this problem demands a very large and computationally impractical number of atoms, so that the energy of the incoming species, after being shared by all atoms in the cell, results in a negligible increase in temperature.

There exist several schemes for dealing with this problem. One of them, by means of a Langevin equation, describes the energy exchanges with the surface through a friction force and a random force with a white noise. [Adelman et al, 1976; Adams et al, 81; Tully et al, 1979; 1980; De Pristo, 1984 ; Brault et al, 1996 ; Guo et al, 2001]. The thermalization occuring during surface diffusion can be accounted for by velocity rescaling [Allen and Tildesley, 1987]. The particle kinetic temperature $T_k$ progressively evolves to the surface temperature $T_s$ in a predefined time interval $\tau$. The velocity lowering factor is given by: $\chi = \left(1 + \frac{dt}{\tau}\left(\frac{T_s}{T_k} - 1\right)\right)^{\frac{1}{2}}$. A more refined model treats the energy exchange as electron-phonon coupling [Flynn et al, 1988; Hou et al, 2000]. The energy exchange is modeled by a friction term and the equations of motion become [Hou et al, 2000]:

$$\frac{\partial^2 \vec{r}_i(t)}{\partial t^2} = -\frac{1}{m_i}\frac{\partial}{\partial \vec{r}}V(\vec{r}_1(t),\vec{r}_2(t),\cdots,\vec{r}_N(t)) - \mu\vec{v}_i(t) \qquad (21)$$

With $\mu = m_i \alpha \frac{T_k - T_s}{T_k}$ and $\alpha = \frac{\Theta_D T_e L n e^2 k_B Z}{2 m_e \kappa \varepsilon_F}$

$\Theta_D$ is the surface Debye temperature, $T_s$ the substrate temperature, L the Lorentz number, n the electron density, e the electron charge, $k_B$ the Boltzmann constant, Z the valence, $m_e$ the electron mass, $\kappa$ the thermal conductivity, $\varepsilon_F$ the Fermi energy and $T_k$ is the kinetic temperature. $1/\alpha$ is the relaxation time.

Finally the most simple way, but not the most rigorous, is to use a simple velocity quenching: As soon as the scalar product $\vec{f}_i(t) \cdot \vec{v}_i(t) < 0$, i.e the incoming particle feels the repulsive part of the interaction potential, the kinetic temperature $T_k = E_k/k_B$ is instantaneously set to the prescribed surface temperature, by randomly selecting velocity components in a Maxwell distribution at $T_s$ [Brault et al, 2002, 2004].

**4. Applications of Molecular Dynamics Simulations to Plasma-Surface Interactions**

The following summaries of MD applied to various plasma-surface or beam-surface studies are meant to be illustrative, and are certainly not comprehensive or exhaustive. The goal is to identify some of the key challenges, successes and opportunities in using MD for interpreting plasma-surface interaction experiments.

4.1 Tetrahedral amorphous carbon film deposition

There are numerous examples of the use of MD to simulate energetic species interacting with carbon surfaces, either from ion beams or plasmas. We will focus here mainly on the formation of tetrahedral amorphous carbon (ta-C) films from energetic (sometimes referred to as 'hyperthermal') C. Other applications involving carbon are addressed in subsequent sections, including plasma-polymer interactions and plasma etch involving carbon either as an etchant or substrate.

ta-C film formation from energetic C (typically 10 eV - 1 keV) has been widely studied, starting apparently with the paper of Aisenberg and Chabot (1971). Lifshitz (1999) reviewed the field relatively recently. Carbon is known to form amorphous networks with threefold-coordinated $sp^2$ bonding and 'glassy carbon,' in addition to ta-C. The latter material consists of a mixture with both threefold-coordinated $sp^2$ bonding and fourfold-coordinated $sp^3$ bonding. In older literature, the material is sometimes referred to as 'diamond-like carbon' (DLC). $Sp^3$ bonding is characteristic of diamond C and imparts various properties leading to hopes of applications including thin film coatings for hard disk drives and cathode materials for field emission displays, among others. Carbon is unique among group IV elements in that it can form amorphous networks containing the threefold-coordinated $sp^2$ hybrid. Si and Ge form fourfold-coordinated ($sp^3$) bonding networks and threefold-coordinated sites are dangling bonds.

The properties of ta-C films formed from energetic C depend sensitively on C impacting energy and surface temperature. It is well established that the fraction of $sp^3$ bonded material is low for impacting energies below about 40 eV, but increases to 60-80% within some energy window, below about 1000 eV. Substrate temperature plays an important role as well. It turns out that above a critical temperature (~ 470 K), the fraction of $sp^3$ bonded material drops dramatically. Other observations include the relation between deposition conditions and cohesive energy, stress, and mass density. Surface roughness is known to be related to $sp^3$ fraction as well. The well-defined nature of the experiments coupled with clear effects associated with the energetic impacting species makes this application a good candidate for study via MD. The question of the mechanisms relating ion bombardment and film stress in ta-C growth, a common issue in plasma-deposited or plasma-exposed thin films, is a topic that has been fruitfully explored by MD. [Abendroth et al., 2007]

Simulating ta-C via MD challenges the method since some of the major potential limitations of MD simulations (accuracy of potentials and accessible simulation length- and time-scales) are known or thought to be critically important to capture experimental observables. Details of the potential can be important for this application since the $sp^3$ bonds are metastable whereas $sp^2$ is the thermodynamically stable form near room temperature and pressure. Certain details of the potential, especially longer-range interactions, are known to be important to faithfully capture these different structures. [McCulloch et al., 2000; Marks, 2002; Jäger and Belov, 2003]

Furthermore, the substrate temperature dependence raises serious questions about time scales.[Marks et al, 2006] This is because the thermally-activated relaxation dynamics transforming $sp^3$ bonds to $sp^2$ bonds may require trajectory simulations for longer times following ion impact than is computationally practical. Questions about length scales in the computational cell arise in discussions of surface roughness. A too-narrow cell cannot capture roughness that develops over distances greater than the cell width. As a result of the combination of the challenges and available, well-characterized experimental data, this application is a good illustration and test of the use of MD to better understand mechanisms of plasma-surface interactions.

Our summary of the published work simulating ta-C film growth relies heavily on the papers of Jäger and Belov [2003], Marks [2002] and Marks et al. [2006]. This set of papers appears to be fairly representative of the field. Jäger and Belov used a Brenner-type empirical potential for the C-C interactions, but with an extended C-C interaction cutoff distance. Marks [2002] discusses the issues in developing an empirical C-C potential that is both 'transferable' and computationally efficient. A transferable potential is one that can faithfully reproduce all important interactions no matter what the relative orientation and distance of the atoms being simulated. In principle, all *'ab-initio'* potentials are transferable, but empirical potential may be less so if their parameters are fit only to, or primarily to, equilibrium configurations.

Figure 1 is a schematic of the simulation cell employed by Jäger and Belov [2003]. A central cylinder of active atoms is surrounded by a region of atoms that are 'heat bathed,' meaning that their energies are re-scaled periodically to maintain a desired temperature using the Berendsen procedure. [Berendsen et al., 1984] Lateral boundaries are periodic and the bottom layers are fixed in place. C ions (assumed to neutralize before impact) are launched at the surface and the trajectory runs for on the order of 10 picoseconds. After this, another impact site is chosen and the process repeats - many thousands of times.

A typical film resulting after 5000 C impacts at 40 eV (normal incidence) onto a room temperature substrate, is shown is shown in Fig. 2. In addition, Fig. 2 plots the depth profile of mass density and symbols showing how C coordination varies with depth. This corresponds to $1.6 \times 10^{17}$ cm$^{-2}$ fluence (flux times time), and resulted in a film that is about 10 nm thick. The key aspects are the initial transition region above the starting diamond (111) substrate, followed by an inner layer with mostly sp$^3$ (4-fold coordinated) carbon, then a surface region that is mostly 3-fold and 2-fold coordinated. The 5-fold coordinated atoms are thought to eventually relax into 4-fold sites. The general picture of a sp$^2$-dominated surface region and a sp$^3$-dominated film deeper in the film is consistent with measurements. Fluctuations in film structure are evident in the 'inner region,' and they are probably due to the relatively narrow central, active region.

This illustrates one potential dilemma of MD simulations: a cell too large will take too long to simulate, whereas one that is too small might introduce too many fluctuations and even artifacts.

Figure 3 shows one of the major results of the paper and a significant success of MD simulations of ta-C growth. The plot shows the percent $sp^3$ bonded C as a function of ion energy and substrate temperature. The lowest temperature substrate (100K) achieves the highest fraction of $sp^3$ C. As substrate temperature is raised, this fraction drops, and between 80C and 130C, the fraction has dropped to less than 20%. This result is semi-quantitatively in agreement with experiment.

Figure 4, however, illustrates some problems with the simulation in the relationship between film mass density and $sp^3$ fraction, especially below about 50% $sp^3$. These values vary between about 2.25 $g/cm^3$ for 0% $sp^3$ (graphite) to about 3.5 $g/cm^3$ for diamond. The MD simulations of Jäger and Belov [2003] strongly over-predict the fraction of $sp^3$ bound C at lower mass density. The EDIP (Environment-Dependent Interaction Potential) developed by Marks [2002] and the density functional theory calculations performed by McCulloch et al. [2000] are in much better agreement with experiment. It has been pointed out by several authors that this problem with the Brenner potential is due primarily to a lack of non-bonded π repulsion in the potential. This defect in the potential leads to $sp^2$ bonded C having un-physically large mass density since the missing repulsive interactions allow atoms with this coordination to come closer to each other than they would in reality. A spurious peak in the radial distribution function just outside the cut-off distance has also been identified for potentials with ad-hoc treatments of the cut-off region interactions. [Marks, 2002] Figure 5 is taken from Marks [2002],

showing generally good agreement between MD simulations of ta-C film deposition using the EDIP and experiment for film stress, $sp^3$ fraction and mass density as a function of incident C energy.

In spite of the difficulties with their potential, Jäger and Belov [2003] were able to develop a deeper understanding of the nature of the film growth process by examining many trajectories to see how C coordination varies in the film during the trajectory. This helped these authors to offer insight into why the $sp^3$ fraction drops above a critical substrate temperature. Figure 6 (Jäger and Belov [2003]) shows results from 500 40 eV $C^+$ impacts onto substrates maintained near 80 °C and 130 °C. Figure 6(a) is the average temperature in the layer over the trajectory. The heat bath allows the cell temperature to rise, but eventually it returns to the set point temperature. The authors note, however, that this unphysical temperature excursion is the likely cause for the simulation predicting that the critical substrate temperature is lower than the experimentally observed value. A wider cell with more effective cooling would be needed, they note, to suppress this effect In Fig. 6(b), curves following the generation of highly coordinated C during the trajectory are presented. The number rises rapidly after impact, but quickly decline as the collision cascade progresses and a process of $sp^3$ to $sp^2$ conversion takes place. The lower temperature layer retains a much higher $sp^3$ fraction at the end of the cascade. Jäger and Belov stress the importance of these relaxation processes occurring after about 0.5 ps following ion impact. Substrate temperatures above the critical value lead to a far greater conversion of $sp^3$ to $sp^2$ during this relaxation period. This general result follows the experimental observation.

We note that Marks et al. [2006] proposed a different scheme to deal with long time-scale relaxation processes governing $sp^3$ to $sp^2$ relaxation. Marks et al. used brief (~1 ps) high temperature pulses between impacts, with the activating temperature ('$T_{act}$') chosen to match the activation energy of the desired infrequent event. Figure 7 shows images of several simulated layers: 7(a) is the original substrate; 7(b) shows the layer after 500 70 eV $C^+$ impacts, using a 1500K temperature pulse between impacts; 7(c) shows the layer using 1875K as the temperature pulse; and 7(d) is a rotated image of 7(c). The $sp^3$ to $sp^2$ relaxation began with $T_{act}$ =1750K, implying an activation energy for this process of about 0.7 eV.

In summary, we believe the successes and challenges in using MD for studies of ta-C deposition illustrate several key points. First, MD simulations (using <u>any</u> potential) must be carefully checked against experiment to ensure that key processes are included properly. The importance of careful comparison between MD predictions and experiment can hardly be overstated. In some cases at least, computationally efficient, empirical, transferable potentials can be developed that are in good agreement with *ab-inito* methods and experiment. Even if the potential has problems, valuable insights can still be obtained from the simulations if intelligently interpreted. Second, although length and time scale limits of MD complicate interpretation of experiments, once again, careful analysis of experimental results with simulation predictions can yield valuable insights that cannot be obtained otherwise. In some cases, schemes can be developed to incorporate infrequent events into MD simulations.

**4.2 Silicon film deposition via radical chemistry**

We turn now to the use of classical MD (and related *ab-initio* tools) to study plasma-surface interactions associated with plasma-enhanced chemical vapor deposition (PECVD} of amorphous hydrogenated silicon (a-Si:H) and nanocrystalline hydrogenated silicon (nc-Si:H) from radical precursors. It turns out that for many such films of interest, including applications to solar cells and thin film transistors, among others, radical-surface interactions are often dominant in controlling film properties under the conditions that produce the highest quality films. This has prompted much study of the mechanisms responsible for the growth rate, structure and composition of the films. For example, as we discuss below, questions involving nanocrystalline domains within the amorphous network have captured considerable interest due to the attractive properties of these films.

The literature that has accumulated on this topic over the past decade or so is arguably the currently most complete atomistic-scale study of chemical details associated with plasma-surface interactions. In this section, we have focused, although not exclusively, on the extensive set of publications of Maroudas and Aydil and their respective teams. These authors have effectively combined MD simulation, *ab-initio* DFT cluster calculations and *in-situ* experimental measurements to identify key mechanisms, and this unusual combination lends weight to their publications. These authors employ a Brenner-type potential with parameterizations for Si-H as originally developed by Ohira. [Ohira et al., 1995; 1996] A good overview of the work of this group in studying PECVD of amorphous Si films is provided by Maroudas. [2001]

4.2.1 SiH$_3$ surface dynamics

In plasmas used to deposit Si:H films, the most common class of precursor gases use the hydrides of silicon, most commonly $SiH_4$, heavily diluted in $H_2$. The radical $SiH_3$ (the 'silyl' radical) has been identified (not without some controversy) as the key Si-containing precursor leading to film deposition. In the next section, we address the very important question of the role of atomic hydrogen (H) in controlling film properties. In this section, we address the work that has been done elucidating the surface reactivity and diffusion of the silyl radical.

Figure 8 illustrates a $SiH_3$ molecule colliding with the surface, abstracting an H to form $SiH_4$, and desorbing. [Ramalingam et al., 1998] This type of prompt reaction involving an impinging reactant from the gas phase is generally referred to as 'Eley-Rideal.' The authors report that the activation energy barrier for this reaction in their simulation is about 0.1 eV. After the H is abstracted, a Si dangling bond is left behind. This is an important reaction because the H-terminated Si surface cannot add more Si.

Although not strictly MD, the images in Figures 9 and 10 illustrate two other characteristic processes occurring between silyl radicals and the hydrogenated Si surface. [Bakos et al., 2006] Both of these processes involve $SiH_3$ migration on the surface. In order to have a more accurate calculation, the MD simulations of Maroudas and colleagues are often supplemented with DFT calculations on clusters of atoms. Generally, similar processes are observed in the MD simulations of amorphous films, but the use of clusters and DFT allows a more detailed analysis of the intrinsic chemical processes. Crystalline silicon forms dimer rows on its 100 surface. In the example chosen for Fig. 9, the silyl radical hops across a broken dimer bond that does not reform after passage of the radical. The barrier is relatively high - about 0.6 eV. If a broken Si-Si bond is reformed, the barrier is generally lower.

Figure 10 illustrates a more complex case, but one that turns out to be important for silyl interacting with a-Si:H surfaces. This figure shows a $SiH_3$ diffusing on the surface by forming a relatively weak, 'over-coordinated' bond with another surface Si that already has four other bonds. Bakos et al. [2005] term this situation 'precursor-mediated' (PM) because the relatively weakly bound silyl is in a 'mobile precursor state.' In the case shown in Fig. 10, this process leads to the $SiH_3$ abstracting an H to form a desorbing $SiH_4$. If the process described in Fig. 9 had led to H abstraction, it would have been termed a 'Langmuir-Hinshelwood' (LH) process. LH processes involve the surface diffusion of one reactant before reaction with another bound reactant to form a product. The difference is in the strength of the surface bonds for the migrating species: LH corresponds to a full surface bond while PM corresponds to a more weakly bound species in a series of hops on over-coordinated surface Si atoms. The typical barriers for LH abstraction are ~0.8 eV and ~0.4-0.5 eV for PM abstraction. [Bakos et al., 2005]

Of course, there are many more reactions and interactions between $SiH_3$ and the a-Si:H surfaces than the ones we highlighted here. The process of film growth, for example, require the Si in the impinging silyl to form a Si-Si bond with one of the surface (or possible sub-surface) Si atoms. An interesting aspect of this reaction is the tendency of the $SiH_3$ to react in surface valleys, promoting surface smoothing. [Valipa et al., 2005] This was explained by a combination of an increased residence time in the valleys for the migrating $SiH_3$ and a decrease in the barriers to reaction.

4.2.2 H surface dynamics

Figures 11 and 12 address another important class of interactions identified by Maroudas et al., namely the role of H atoms at surfaces. The figures illustrate the role of diffusing H in creating nano-crystalline domains in the nominally hydrogenated amorphous silicon network. It has been known experimentally for many years that H can create crystalline domains, but the mechanisms were speculative. The MD simulations, summarized in Figs. 11 and 12 illustrate one possible mechanism.

The basic idea is that H diffusing through the film can insert temporarily as a bridge between strained Si-Si bonds. When the H moves away, the strained bonds can relax in concerted moves involving multiple atoms. This often leads to the Si atoms being able to move into lower energy, crystalline configurations. Sriraman [2002] et al. refer to this as a 'chemically induced ordering mechanism.'

Figure 11 shows part of an amorphous hydrogenated Si film (H not shown for clarity) before and after H exposure. [Sriraman et al., 2002] Crystalline Si is also shown in the third column for comparison. Without worrying about the precise mechanistic details, which are not easy to discern from the figure, one can see clearly that the post-H exposure material more closely resembles crystalline Si.

Figure 12 illustrates one way the process can occur. [Valipa, 2006] A diffusing H moves into a bond center (BC) configuration (seen in the image 'TS' in (c)) between two Si atoms that had not been bound before. The energetics of the reaction along the reaction coordinate is shown in 12(a). The barrier is about 0.4 eV, and the final configuration is about 0.2 eV more stable than the starting configuration. Figure 12(e) shows the Si-Si distance during the approximately 1.5 ps encounter - they end up 2.5 Å apart after starting with a separation of about 3 Å. Figure 12(f) shows Si-H distances. This is only one of many ways that this mechanism can occur, of course. But once again, the striking importance of atoms (in this case H and Si) in over-coordinated configurations, leading to diffusion and reaction, is noteworthy. We note in passing that it was important to establish that the modified Ohira-Brenner potential used is capable of reasonable accuracy for these over-coordinated interactions.

4.2.3 Si nano-particles

We end this section with a brief discussion of representative MD calculations on Si nano-particles. Silicon nano-particles form in the gas phase or in a low temperature plasma and grow through aggregation or accretion of silicon-containing radicals. H adsorption on the surface and incorporation into the particle alter growth and aggregation dynamics. Figure 13 illustrates one of the calculations made by Hawa and Zachariah [2005] . They used an empirical potential for the Si-Si, Si-H and H-H interactions and performed classical MD simulations. These authors show how H-coated or non-coated 6 nm diameter Si particles sinter over a period of about 150 ps. The reduced moment of inertia, shown for the two cases, is a convenient measure of the degree to which the particles have fully merged into one. The effects of H coating can be clearly seen in slowing this kinetics of the process.

Vach and Brulin also address the question of silicon nanoparticles, but they examined much smaller particles and used a semi-empirical PM3 method that they term 'quantum MD.' [Vach and Brulin, 2005] They also took advantage of a lower level plasma fluid model to establish the dominant reactive species and their energies in the gas phase. The MD simulations were run to examine primarily the effect of different concentrations (equivalently, fluxes) of H on the growing Si nanoparticles. Part of these results are reproduced as Fig. 14. Vach and Brulin emphasize the importance of H atom fluxes in controlling the growing particle structure. Counter-intuitively, they show that lower fluxes of H atoms lead to particles with higher H concentration and vice-versa. This is due to higher particle temperature for the case of H flux, resulting in H desorption. Both H adsorption into Si dangling bonds at the surface and abstractive recombination of H to form $H_2$ cause significant particle heating. This heating allows particles to form different configurations when finally cooled.

4.3 Plasma Etching

The application of MD to studies related to plasma etching (including chemistry) began with the development of a transferable many-body potential to the Si-F system by Stillinger and Weber. [1989]. Since this study, many papers have appeared, and we don't attempt to list them all. [see, e.g., Graves, 2001] A recent review by Gou, Kleyn and Gleeson [2008] of MD simulations of $CF_x$ interacting with Si surfaces has many useful details and references. More complete lists of references can also be found in PhD theses. [Barone, 1995]; [Helmer, 1998]; [Abrams, 2000]; [Humbird, 2004]; Vegh [2007].

4.3.1 Spontaneous etching

The term 'spontaneous etching' in the context of plasma etching generally refers to etching in the absence of energetic ion bombardment from neutral species. As noted above, MD studies of 'plasma' etching began with the development of the Stillinger-Weber (SW) potential for the Si-F system. Naturally, the initial studies consisted of trying to simulate the removal of silicon by F atoms at room temperature, which is well known experimentally to result in Si etching. [Winters and Coburn, 1992] Unfortunately, none of the studies using the SW potential predicted etching with 300K F atoms impacting the 300K Si surface. F uptake to the Si surface stopped after a little over one monolayer of F chemisorbed. [Stillinger and Weber, 1989; Schoolcraft and Garrison, 1991; Weakliem et al., 1992; Weakliem and Carter, 1993] studied Si etching by employing 3 eV F directed to the surface. In the absence of several eV kinetic energy directed towards the surface, or by using a mixture of rare gas ion bombardment and F atoms, no etching was observed. This qualitative, and fundamental, disagreement with experiment was explained in part by the known limitations of the potential.

To address this problem, Weakliem et al. [1992] and Weakliem and Carter [1993] used *ab-initio* quantum calculations to improve the SW Si-F potential. However, the changes they suggested to the parameters in the potential did not result in MD simulations that predicted actual Si removal from the surface. One possible explanation is that spontaneous etching of Si by F is not prompt, requiring a longer surface residence time than can be followed in MD. Another possibility is that the potential itself was still significantly in error for some rate-limiting aspects of the process, in spite of the improvements made by Weakliem et al. Later work strongly suggests that the latter explanation was the correct one.

Figure 15 is a sketch of a potential energy profile along a trial reaction coordinate illustrated in the figure of an F atom approaching a fluorinated Si cluster, resulting in F insertion into the Si-Si bond. The plot using the symbols denoted 'potential' were from a calculation using the SW potential, and the symbols 'DFT' correspond to the cluster calculations (using density functional theory) of Walch. [2002] Clearly, the DFT calculations show no barrier for this insertion process, whereas the SW potential predicts a barrier of about 3 eV. Apparently, the SW potential introduces some spurious barriers for some trajectories of F approaching a fluorinated Si surface. This result is consistent with the calculations of Schoolcraft and Garrison that resulted in Si etching if the impinging F had 3 eV.

Humbird and Graves [2004a,b] used a modified Tersoff-Brenner potential for Si-F and Si-Cl interactions to overcome the aforementioned problems with the Stillinger-Weber form. Their revised potential was based on earlier work. [Abrams and Graves, 1999] In essence, Humbird and Graves used the ab-initio cluster calculations of Walch [2002] to re-parameterize the Abrams-Graves potential. Figure 16 illustrates one representative trajectory, using this potential, for the formation of $SiF_4$ from a surface $SiF_3$ species reacting promptly with an incident F atom. Simulations were able to reproduce several important experimental observables, including steady state F-Si etch reaction probabilities; etch product distribution; and the change in etch product from $SiF_4$ and $Si_2F_6$ to $SiF_2$ as surface temperature increased. The simulations under-predicted etch rates as surface temperature was raised, probably because the dominant etch mechanism shifted from direct abstraction to $SiF_x$ layer decomposition as temperature increased. The latter mechanism requires considerably longer trajectory simulations, and it is likely that the MD simulations of less than 5 ps trajectories failed to capture these longer-timescale etch events.

4.3.2 Ion-assisted etching

There have been many studies of ion-assisted etching with MD and we do not attempt to cite the entire literature. More complete reference lists can be found in Abrams [2000], Schoolcraft [2001], Humbird [2004], Vegh [2007] and Gou et al. [2008]. We have chosen, rather, to highlight several studies that have focused on some of the key problems associated with ion-assisted etching. The previous section presented results from spontaneous etching reactions - that is, etching that resulted from only radical impact at room temperature. When energetic ions are added to the mix, the effects can be quite different. [Winters and Coburn, 1992] The energy deposition and mixing induced by ion bombardment alters the surface chemistry profoundly. As Graves and Humbird [2002] point out, ion impacts in plasma etching are isolated in time and space. These isolated impacts deliver enormous peak power to very small volumes at the surface over very short times, but relatively little average power. The highly localized peak power breaks chemical bonds and promotes mixing, resulting in dramatic effects on surface composition and structure. But the low average power keeps the material from over-heating and vaporizing. The plasma combines energy from ion impact with chemical effects from near room temperature radicals to achieve unique effects in surface chemical processing. Although we do not discuss it further, it is no doubt true that plasma-generated electrons and photons also play important roles in surface chemistry under some conditions.

The first illustration of the effects of combined ions and radicals in shown in Fig. 17. The problem addressed is fluorocarbon (FC) plasma etching of silicon. In this case, no FC film has formed on the surface (discussed next), but under conditions of steady state etching, an unexpected effect was observed: Si-C and Si-F layers formed spontaneously. The Si-F layer is the leading front for etching and the Si-C layer inhibits etching considerably. The sketch in Fig. 16 shows how $Ar^+$ bombardment at different depths promotes the formation of SiF etch products near the leading edge, followed by transport of this Si into the Si-C layer, and finally the formation of a silicon fluoride etch product at the surface. All of the processes happen randomly at a given location, depending on how the ion penetrates the layer.

It has been widely documented experimentally that a FC film often forms on the surface when FC etch gases are used. [Oehrlein et al., 1994] The nature of this FC film and the role it plays in inhibiting etching has remained controversial. The MD simulations reported by Vegh et al. [2005] help to shed some light on the mechanisms. These authors found that if CF or $C_4F_4$ were used as the fluorocarbon film deposition precursors, along with F atoms to act as primary etchant and $Ar^+$ to supply the needed energy, then steady state FC films with steady etching of the underlying silicon would be observed in the etch simulations. One sequence of images showing the formation of the FC layer (on top of the previously-noted Si-C and Si-F layers) is presented in Fig. 18. The key result is that for FC films to form during steady etching, they must be porous and 'fluffy.' The simulation shows that these films fluctuate locally, depending on how the FC film precursors randomly deposit and how the ions and F atoms promote etching. The FC film is constantly etching and re-depositing, resulting in some average film thickness, while allowing the underlying Si to be etched. If the film is too thick or dense, ion bombardment and radical attack of the underlying film cannot take place. Figure 19 shows the simulation result of the effect of the average film thickness on average etch yield, following the trends reported experimentally.

The question of film formation, surface modification and etching was addressed by Kawase and Hamaguchi [2007] for an important practical etch application: FC plasma etching of $SiO_2$. Typical results from one of their MD simulations is illustrated in Fig. 20. The images show layer side-views after ion fluences of $2 \times 10^{16}$ cm$^{-2}$ at two energies ((a) 300 eV and (b) 200 eV) for $CF_3$ beams impacting $SiO_2$ at normal incidence. At the higher energy, the layer continuously etches, but no FC film has formed. A mixed top layer is seen, consisting of Si, C, O and F. At the lower energy, however, a FC film deposits continuously with no etching of the underlying substrate.

A similar study was reported by Smirnov et al. [2007] for an important low dielectric constant ('low-κ') material, containing Si, C, O and H ('SiCOH'). These authors found that a FC passivation layer forms in the near-surface region when impacting their model dielectric film with $CF_x$ ions. Figure 21 illustrates the structure from one of their porous film etching simulations. Theses authors noted that film porosity can play an important role in the nature of the etch process. Atomic, molecular and cluster-like structures were observed to be sputtered from the surface.

We end the presentation of typical MD studies of plasma etching with a result from a relatively old publication, but one that addresses a topic that continues to draw interest: namely, atomic layer etching. One of the mechanisms of $Ar^+$ ion-assisted Cl etching of Si is shown in Fig. 22. [Kubota and Economou [1998] Atomic layer etching - meaning that the surface is etched one atomic layer at a time - continues to be of practical importance because it is increasingly desirable and even necessary to etch structure with nano-scale precision. If this can be done in a controllable way, many new electronic, optical and other devices may be built economically using plasma etching. Figure 22 shows how an impacting $Ar^+$ can cause an adsorbed Cl to move to a surface SiCl moiety, creating a $SiCl_2$ that desorbs promptly.

## 5. Concluding Remarks

MD is a powerful tool to understand the mechanistic details of interactions between plasma-generated species and surfaces. The primary limitations of MD are the relatively small number of atoms that can be simulated for relatively short times and the approximations associated with inter-atomic potentials. It turns out that many important questions surrounding plasma-surface interactions can be addressed in spite of these limitations. Although we did not attempt to address the literature in this topical review, developments are under way to extend and augment MD with other methods in order to address longer time and length scales, as well as to improve the accuracy of the inter-atomic potentials.

It was stressed that MD is often best utilized in the context of interpreting experimental observations. The uncertainties in the approximations and assumptions that are commonly involved in simulating plasma-surface interactions are best dealt with by direct comparisons between predictions and well-defined measurements. The applications of MD to plasma-surface studies that are highlighted in this article are examples of this principle. MD simulations of energetic C impacting on a growing ta-C film were able to reproduce important aspects of the process, albeit sometimes qualitatively. The complex processes occurring in the top several nanometers and during the several picoseconds following energetic species impact of the surface were identified using MD, although many questions remain. Similar comments can be made regarding the interactions of radicals with amorphous hydrogenated silicon films. For example, the role of H atoms at surfaces in causing strained Si-Si bonds to relax into crystalline orientations resulted from a combination of experimental observations and MD simulations. Ion-assisted and spontaneous etching simulations were first validated or at least supported by comparisons to measurements, then the simulations provided important details of mechanisms.

Atomistic simulation is now established as a powerful complementary tool in scientific experimental and modeling investigations of complex phenomena in technology and nature. Increasingly powerful computational platforms, algorithms and simulation strategies should augment this trend in the future. We anticipate that these advances will continue to help shed light on the complexities of low temperature plasmas altering surfaces.

## 6. Acknowledgements

DBG gratefully acknowledges the support of le STUDIUM and GREMI during the preparation of much of this paper. He also thanks the students, postdoctoral scholars and visitors who have helped teach him about the use of MD for plasma-surface interactions.

Table 1. Pair potential parameters corresponding to Eq. (1, 2 and 3) for various species (from Halicioglu et al, 1975; Girifalco et al, 1959; Flahive et al., 1980; Chisholm et al, 1999.

| species | Morse potential | | | L-J potential | | Buckingham potential | | |
|---|---|---|---|---|---|---|---|---|
| | $D_0$ (eV) | $\alpha$ (Å$^{-1}$) | $r_0$ (Å) | $\varepsilon$ (Å) | $\sigma$ (eV) | A (eV) | $r_B$ (Å) | $C_6$ (eV Å$^6$) |
| Ag | 0.3294 | 1.3939 | 3.096 | | | | | |
| Al | | | | 0.392 | 2.62 | 5131.179 | 0.3224 | 248.0 |
| Ar | | | --- | 0.01 | 3.4 | | | |
| Au | 0.4826 | 1.6166 | 3.004 | 0.449 | 2.637 | | | |
| Ba | 0.1416 | 0.65698 | 5.373 | | | | | |
| C | | | | 3.4 | 2.41 10$^{-3}$ | | | |
| Ca | 0.1623 | 0.80535 | 4.569 | 0.215 | 3.6 | | | |
| Cr | 0.4414 | 1.5721 | 2.754 | 0.502 | 2.336 | | | |
| Cs | 0.04485 | 0.41569 | 7.557 | | | | | |
| Cu | 0.3446 | 1.3921 | 2.864 | 0.409 | 2.338 | | | |
| Fe | 0.4216 | 1.3765 | 2.849 | 0.527 | 2.321 | | | |
| Ga | --- | -- | | | | 5902.871 | 0.3187 | 250.0 |
| He | | | | 8.81 10$^{-4}$ | 2.56 | | | |
| In | --- | -- | | | | 6141.774 | 0.3567 | 258.0 |
| Ir | 0.8435 | 1.6260 | 2.864 | | | | | |
| K | 0.05424 | .49767 | 6.369 | 0.114 | 4.285 | | | |
| Kr | | | | 0.014 | 3.65 | | | |
| Li | | | | 0.205 | 2.839 | | | |
| Mo | 0.7714 | 1.434 | 3.012 | 0.838 | 2.551 | | | |
| N | 10.56 | 2.557 | 1.097 | | | 5134.176 | 3.140 | 283.8 |
| Na | 0.06334 | 0.58993 | 5.336 | 0.1379 | 3.475 | | | |
| Nb | 0.9437 | 1.5501 | 3.079 | | | | | |
| Ne | | | | 3.13 10$^{-3}$ | 2.74 | | | |
| Ni | 0.4279 | 1.3917 | 2.793 | 0.520 | 2.282 | | | |
| O | 5.12 | 2.68 | 1.208 | | | | | |
| Pb | 0.2455 | 1.2624 | 3.667 | 0.236 | 3.197 | | | |
| Pd | 0.4761 | 1.6189 | 2.89 | 0.427 | 2.52 | | | |
| Pt | 0.7102 | 1.6047 | 2.897 | 0.685 | 2.542 | | | |
| Rb | 0.04644 | 0.42981 | 7.207 | | | | | |
| Rh | 0.6674 | 1.5423 | 2.875 | | | | | |
| Sr | 0.1513 | 0.73776 | 4.988 | | | | | |
| W | 0.9710 | 1.385 | 3.053 | 1.068 | 2.562 | | | |
| Xe | | | | 0.02 | 3.98 | | | |
| AlN | | | | | | 698.647 | 0.3224 | 0. |
| GaN | | | | | | 782.107 | 0.3166 | 0. |
| InN | | | | | | 870.207 | 0.3263 | 0. |

Table 2: Parameters of Moliere and ZBL potentials, Bourque et al, 1998

| i | $c_i$ | $d_i$ |
|---|---|---|
| | Moliere | |
| 1 | 0.35 | 0.3 |
| 2 | 0.55 | 1.2 |
| 3 | 0.1 | 6.0 |
| | ZBL | |
| 1 | 0.02817 | 0.20162 |
| 2 | 0.28022 | 0.40290 |
| 3 | 0.50986 | 0.94229 |
| 4 | 0.18175 | 3.19980 |

Table 3: SiO$_2$ parameters for BKS and Garofalini-Feuston potentials, Levine et al, 1986

| i − j | $A_{ij}$ (eV) | $b_{ij}$ (Å$^{-1}$) | $C_{ij}$ (eV Å 6) | charges q |
|---|---|---|---|---|
| | BKS | | | |
| O -O | 1388.7730 | 2.76000 | 175.0000 | $q_O$ = -1.2 |
| Si -O | 180003.7572 | 4.87318 | 133.5381 | $q_{Si}$ = 2.4 |
| i − j | $B_{ij}$ (eV) | $\beta_{ij}$ Å) | $r_G$ (Å) | charges q |
| | Garofalini | | | |
| O -O | 452.53 | 2. 34 | 0.29 | $q_O$ = -2 |
| Si -O | 1848.8 | 2.34 | 0.29 | |
| Si-Si | 1171.6 | 2.30 | 0.29 | $q_{Si}$ = 4 |

Table 4: Parameters of Tersoff and Brenner potentials (Brenner 1990; Tersoff 1988a,1988b, Albe 1998).

|  | Tersoff (Si) | Tersoff (C) | Brenner (C) | BN | NN | BB |
|---|---|---|---|---|---|---|
| $D_0$ (eV) | 2.666 | 5.1644 | 6.325 | 6.36 | 9.91 | 3.08 |
| $R_0$ (Å) | 2.295 | 1.447 | 1.315 | 1.33 | 1.11 | 1.59 |
| $S$ | 1.4316 | 1.5769 | 1.29 | 1.0769 | 1.0769 | 1.0769 |
| $\beta$(Å$^{-1}$) | 1.4656 | 1.9640 | 1.5 | 2.043057 | 1.927871 | 1.5244506 |
| $\gamma$ | $1.1\ 10^{-6}$ | $1.5724\ 10^{-7}$ | $1.1304\ 10^{-2}$ | $1.1134\ 10^{-5}$ | 0.019251 | $1.6\ 10^{-6}$ |
| $n$ | 0.78734 | 0.72751 | 1 (1/2n = 0.80469) | 0.364153367 | 0.6184432 | 3.9929061 |
| $c$ | $1.0039\ 10^5$ | $3.8049\ 10^4$ | 19 | 1092.9287 | 17.7959 | 0.52629 |
| $d$ | 16.217 | 4.384 | 2.5 | 12.38 | 5.9484 | 0.001587 |
| $h$ | -0.59825 | -0.57058 | -1 | 0.5413 | 0 | 0.5 |
| $\lambda$ | 0 | 0 | 0 | 1.9925 | 0 | 0 |
| $R$ (Å) | 2.85 | 1.95 | 1.85 | 2.0 | 2.0 | 2.0 |
| $D$ (Å) | 0.15 | 0.15 | 0.15 | 0.1 | 0.1 | 0.1 |

Table 5: Parameters of Vashishta potential for $SiO_2$ and $Si_3N_4$ (Vashishta et al, 1997)

| | $A_{ij}$(erg) | $r_{1s}$ (Å) | $r_{4s}$ (Å) | $r_c$ (Å) | $\ell$ (Å) | $r_{c3}$ (Å) |
|---|---|---|---|---|---|---|
| SiO2 | $1.242 \cdot 10^{-12}$ | 4.43 | 2.5 | 5.5 | 1.0 | 2.6 |
| $Si_3N_4$ | $2.00 \cdot 10^{-12}$ | 2.5 | 2.5 | 5.5 | 1.0 | 2.6 |

| | $\sigma_i$ (Å) | $Z_i$ (e) | $\alpha_i$ (Å$^3$) |
|---|---|---|---|
| Si | 0.47 | 1.20 | 0.00 |
| O  | 1.20 | -0.60 | 2.40 |
| Si | 0.47 | 1.472 | 0.00 |
| N  | 1.30 | -1.104 | 3.00 |

| | $\eta_{ij}$ | | $B_{jik}$(erg) | $\theta_{jik}$(°) |
|---|---|---|---|---|
| Si-Si | 11 | Si-O-Si | $3.20 \cdot 10^{-11}$ | 141.00 |
| Si-O  | 9  | O-Si-O  | $0.80 \cdot 10^{-11}$ | 109.47 |
| O-O   | 7  | | | |
| Si-Si | 11 | Si-N-Si | $2.0 \cdot 10^{-11}$ | 120.00 |
| Si-N  | 9  | N-Si-N  | $1.0 \cdot 10^{-11}$ | 109.47 |
| N-N   | 7  | | | |

Table 6: Parameters of TB-SMA potential for cfc metals (Z=12) following Rosato et al 1989. Parameters are deduced using fitting up to 1$^{st}$ neighbor. A and $\xi$ can be deduced using Equation (11).

|        | Ni    | Cu    | Rh    | Pd    | Ag    | Ir    | Pt    | Au    |
|--------|-------|-------|-------|-------|-------|-------|-------|-------|
| a (Å)  | 3.52  | 3.61  | 3.80  | 3.89  | 4.09  | 3.84  | 3.92  | 4.08  |
| $E_c$ (eV) | 4.44  | 3.50  | 5.75  | 3.94  | 2.96  | 6.93  | 5.86  | 3.78  |
| p      | 10.00 | 10.08 | 14.92 | 10.84 | 10.12 | 14.53 | 10.80 | 10.15 |
| q      | 2.70  | 2.56  | 2.51  | 3.67  | 3.37  | 2.90  | 3.50  | 4.13  |

Table 7: Parameters of TB-SMA potential for cfc metals and Al, Pb following Cleri et al 1993. All parameters A, ξ, p, q are deduced using fitting up to $5^{st}$ neighbor.

|   | Ni     | Cu     | Rh     | Pd     | Ag     | Ir     | Pt     | Au     | Al     | Pb     |
|---|--------|--------|--------|--------|--------|--------|--------|--------|--------|--------|
| A | 0.0376 | 0.0855 | 0.0629 | 0.1746 | 0.1028 | 0.1156 | 0.2975 | 0.2061 | 0.1221 | 0.0980 |
| ξ | 1.070  | 1.224  | 1.660  | 1.718  | 1.178  | 2.289  | 2.695  | 1.790  | 1.316  | 0.914  |
| p | 16.999 | 10.08  | 18.45  | 10.867 | 10.928 | 16.980 | 10.612 | 10.229 | 8.612  | 9.576  |
| q | 1.189  | 2.56   | 1.867  | 3.742  | 3.139  | 2.961  | 4.004  | 4.036  | 2.516  | 3.648  |

**Figure Captions.**

Fig. 1 Schematic of MD cell used to simulate ta-C deposition from energetic $C^+$ impact. Atoms in central cylinder move with exact potentials; atoms in outer region are 'heat bathed.' Atoms in bottom region are fixed. [Jäger and Belov, 2003)

Fig. 2 Depth profile of mass density and C coordination, and sideview image of ta-C deposition after an ion fluence of $1.6 \times 10^{17}$ cm$^{-2}$, or 5000 40 eV $C^+$ on this cell. Note the transition region above the diamond substrate, followed by an inner film region with mostly $sp^3$ 4-fold coordination and surface region with mostly $sp^2$ 3-fold coordination. Significant statistical fluctuations are seen in this relatively narrow cell. [Jäger and Belov, 2003)

Fig. 3 Plot of ta-C film $sp^3$ percent (averaged from MD simulations) vs. ion energy for different substrate temperatures. At 40 eV $C^+$ and at the lowest temperature (100K), the percentage of $sp^3$ bonded C rises to about 80%. Higher substrate temperature causes the percentage to drop. Between 80C and 130C, $sp^3$ content drops precipitously. This trend is in semi-quantitative agreement with experiment. [Jäger and Belov, 2003)

Fig. 4 Plot of simulated film mass density vs. $sp^3$ percent from experiment and from other theoretical approaches. The low $sp^3$ percent region results using the modified Brenner potential employed by Jäger and Belov are seen to be in significant disagreement with experiment and the more accurate EDIP simulations of Marks(2002) and density functional theory (DFT) results of McCulloch et al. ( [Jäger and Belov, 2003)

Fig. 5 Plot of experiment and simulation predictions of film stress, $sp^3$ percent and density, using the EDIP potential developed by Marks. [Marks, 2002]

Fig. 6 Comparison of cell properties, averaged during 500 40 eV $C^+$ trajectories, for a cell maintained at 130C and 80C. The time-dependence illustrates the essentially dynamic character of creation and loss of highly coordinated C in the film. (a) Average cell temperature vs. time; (b) Number of 4-fold or 5-fold coordinated C generated during trajectory. [Jäger and Belov, 2003)

Fig. 7 Illustration of the use of high temperature pulsing between impacts to simulate activated, infrequent events for the case of ta-C. 500 70 eV $C^+$ impacts were followed on a 800K substrate (a). Images in (b) and (c) were obtained using 1500K and 1875K temperature pulses between impact. Image (d) in (c) is rotated. Color indicates coordination: red (sp), green ($sp^2$) and blue ($sp^3$). [Marks et al., 2006]

Fig. 8 Three configurations showing how $SiH_3$ interacts with H-terminated surface, resulting in abstraction of surface bound H, creating $SiH_4$ that desorbs back into the gas phase. This direct abstraction mechanism is Eley-Rideal. [Ramalingan et al., 1998]

Fig. 9 Schematic from DFT calculation showing migration of SiH$_3$ across dimer rows on Si(001)-(2x1):H. Selected Si (dark) and H (white) spheres are shown. Si1 is in the migrating SiH$_3$. Migration occurs (a-f) as Si1 breaks bond with Si2 and forms bridge bond with Si3, then transfer completely to Si3. Color contour shows valence electron density (blue: high; red: low density); (g) is total energy map showing activation energy barrier along the migration reaction coordinate [Bakos et al., 2006]

Fig. 10 Mechanisms of H abstraction on Si(001)-(2x1):H surface by migrating SiH$_3$ in the precursor-mediated (PM) mechanism (a-f). PM mechanism involves SiH$_3$ forming weak over-coordinated Si-Si bonds while hopping from site to site, before abstracting an H from one of the surface Si atoms. Dotted lines indicate weak bonds. Numbers indicate inter-atomic distances in Å; VED map for each configuration shown below (g-l). [Bakos et al., 2005]

Fig. 11 Simulated structural characteristics of Si film (H suppressed for clarity) before (a, d) and after (b, e) H exposure, from the two indicated directions. Local structural rearrangements and configuration shown in g and h, respectively. Si-Si bonds under tensile and compressive strain indicated by red and blue, respectively. Corresponding Si crystalline clusters shown in c, f and i for comparison. [Sriraman et al., 1998]

Fig. 12 (a) Energetics along reaction path for H atom diffusing into a bond center (BC) location between two initially un-bonded Si atoms. The Si-Si bond forms after the H leaves the BC site. (b)-(d) show structures for the initial (A, b) to final states (B, d), through the transition state (TS, c) (e) shows Si-Si distance during the simulated trajectory; (f) are the Si-H distances participating. [Valipa et al., 2006]

Fig. 13 Temporal snapshots of the morphology during sintering of H-coated 6 nm Si particles. Inset graphs show temporal behavior of the reduced moments of inertia for coated and bare particles. The coated particle sintering shows a delay. [Howa and Zachariah, 2005]

Fig. 14 (a) Typical example for an amorphous nanoparticle structure created from growth in a pure silane plasma, aggregation of which can lead to dust; (b) typical structure of H-rich, crystalline Si nanoparticle resulting from low H flux conditions; (c) typical structure of H-poor, crystalline Si resulting from high H flux conditions; (d) side and top views of typical tube-like structures that result from intermediate H flux conditions. [Vach and Brulin, 2005]

Fig. 15 Energetics along reaction path for F atom inserting into Si-Si bond with F chemisorbed on the Si atoms. The squares ('potential') are from a calculation using the Stillinger-Weber potential and the solid diamonds are from the DFT calculations of Walch. [Humbird and Graves, 2004a]

Fig. 16 Schematic illustration of an important mechanism for spontaneous etching of Si by thermal (300K) F atoms. (a) F approaching a surface –SiF3 group; (b) transition state; and (c) Formation and desorption of volatile $SiF_4$ product. [Humbird and Graves, 2004b]

Fig. 17 Side-view schematic illustration of the events leading to etching of a Si layer under state state conditions, with fluorocarbon etchants and argon ions. In this case, no fluorocarbon film forms, but ~ nm thick layers comprising predominately Si-C and Si-F form spontaneously. (a) Ion bombardment in the SiF layer causes F to mix into the Si region, advancing the SiF front; (b) ion impacts in the SiF layer help SiF2 to move to the Si-C layer; (c) Si attaches to C in the Si-C layer; (d) ion bombardment in the Si-C layer helps the Si diffuse through the layer; (e) Si moves to the top of the surface and a new Si moves into the Si-F layer; (f) gas phase incident F reacts with near-surface Si; (g) near-surface ion bombardment promotes product release. All of these events are occurring randomly depending on the depth of individual ion bombardment events, and layer depths and composition will fluctuate from point to point. [Humbird and Graves, 2004c]

Fig. 18 Side-view illustration of Si etch with fluorocarbon neutrals and argon ions under conditions that a fluorocarbon film forms on the surface. (a) initial Si layer; (b) steady state layer during etch, showing fluorocarbon (FC) film above the Si-C and Si-F layers shown in Fig. 17; (c) image following one ion trajectory that reduced FC film thickness locally; (d) FC film has built up again. This sequence illustrates the fluctuating nature of the FC film at the surface during etch through a FC film. [Vegh, 2007]

Fig. 19 Plot of average steady state Si etch yields vs. average FC film thickness for impacts with 200 eV $Ar^+$ and CF or $C_4F_4$ and F. Similar behavior has been observed experimentally by many authors. [Vegh, 2007]

Fig. 20 Sideviews of structure after fluences of 2 x $10^{16}$ $cm^{-2}$ at (a) 300 eV and (b) 200 eV $CF_3$ beam impacting $SiO_2$ at normal incidence. In (a), etching takes place, and a mixed layer of Si, C, O and F forms. Steady state $CF_x$ film deposition occurs in (b). Atomic composition depth profiles shown at the sides. [Kawase and Hamaguchi, 2007]

Fig. 21 Sideview of porous low dielectric constant SiCOH film showing passivation layer under etching conditions. [Smirnov et al., 2007]

Fig. 22 Snapshots of simulated Ar ion impacts promoting adsorbed Cl etching Si (a) Cl recoils across the dimer channel (b) and forms a bond with a Si atom (c), resulting in $SiCl_2$ formation and desorption. [Kubota and Economou, 1998]

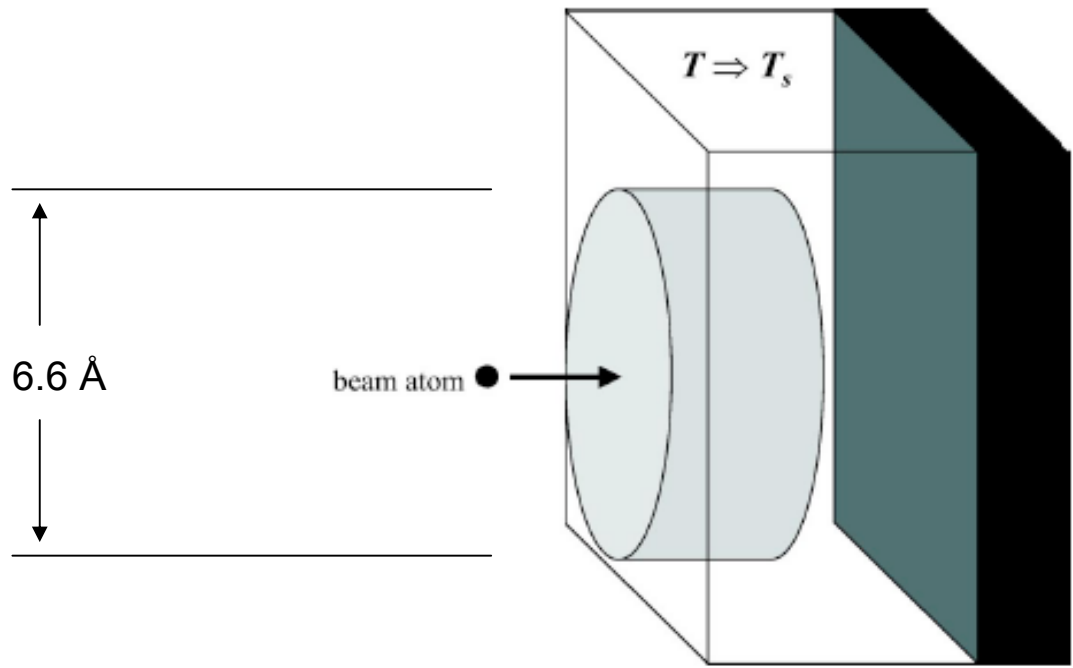

Figure 1. Graves and Brault

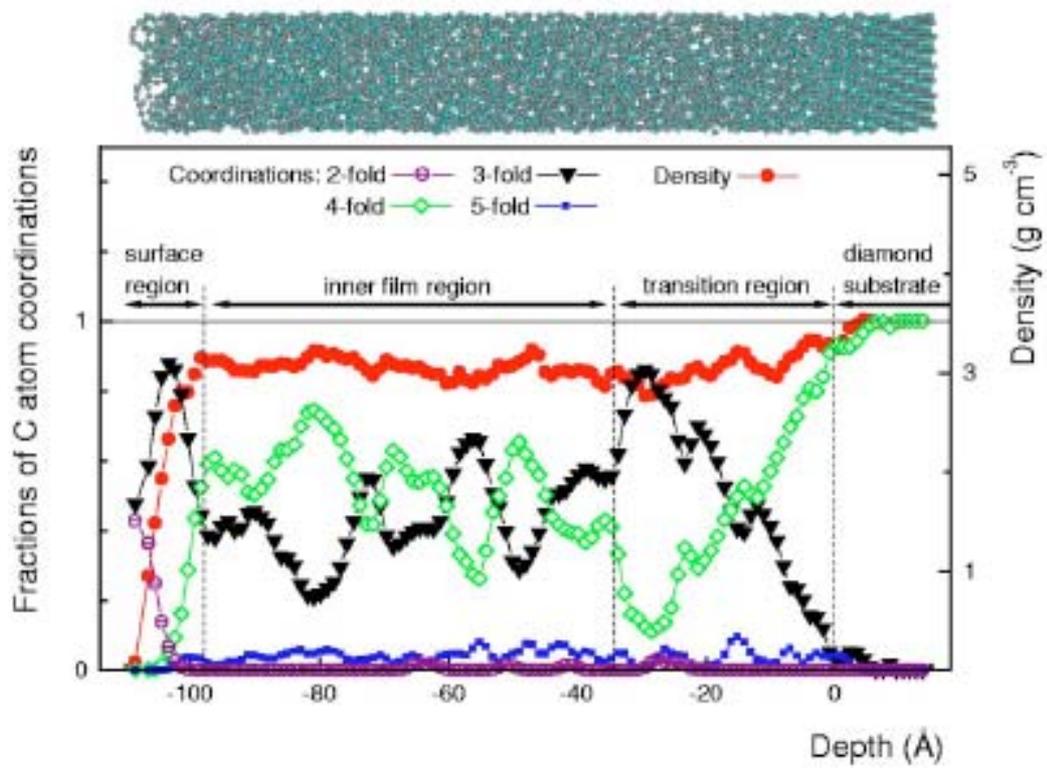

Figure 2. Graves and Brault

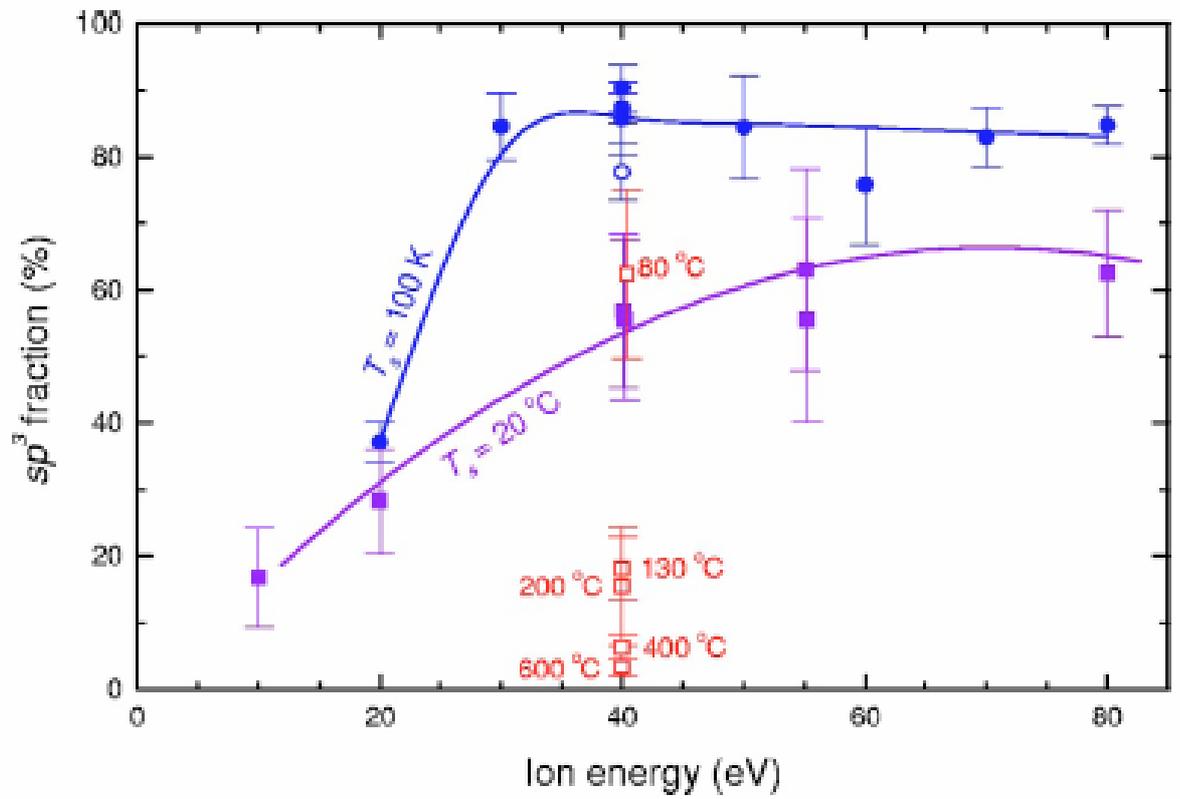

Figure 3. Graves and Brault

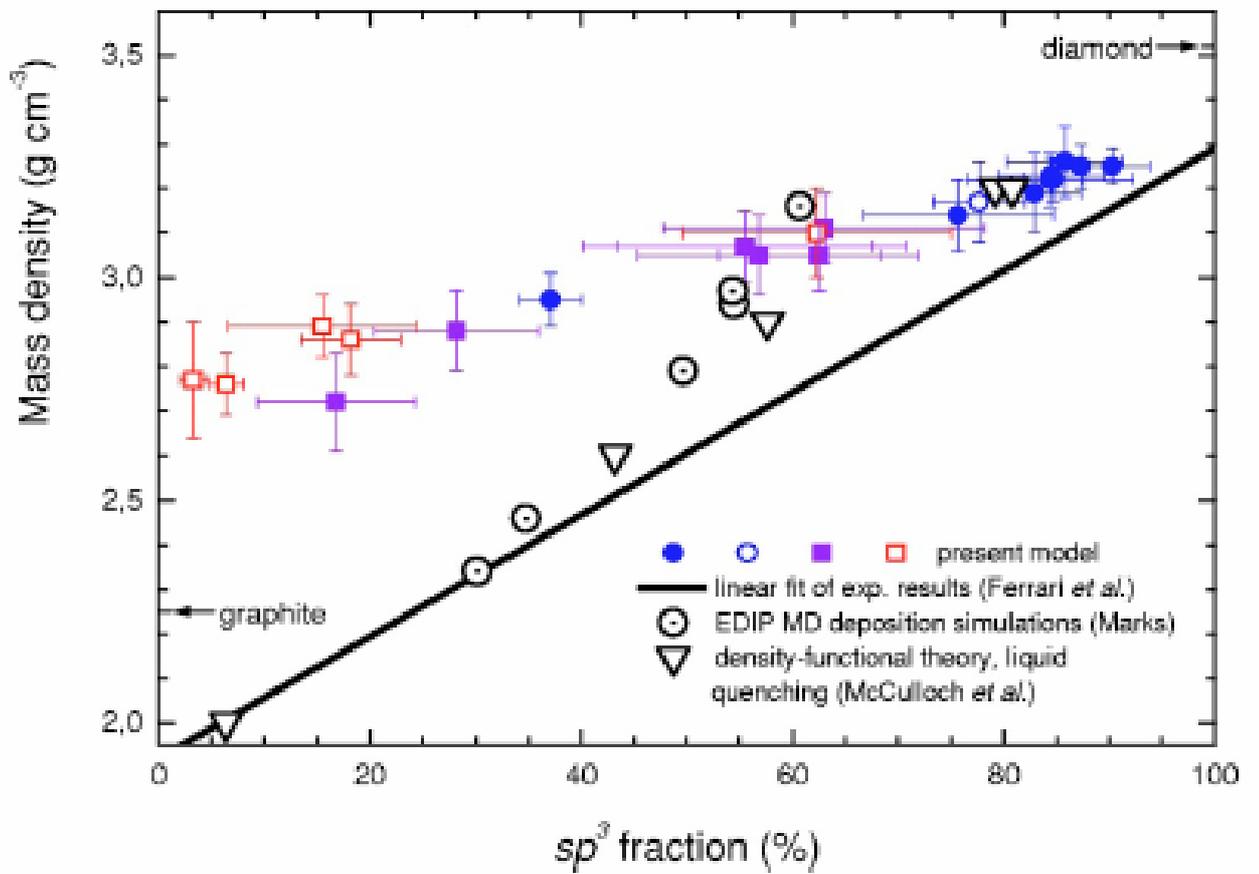

Fig. 4  Graves and Brault

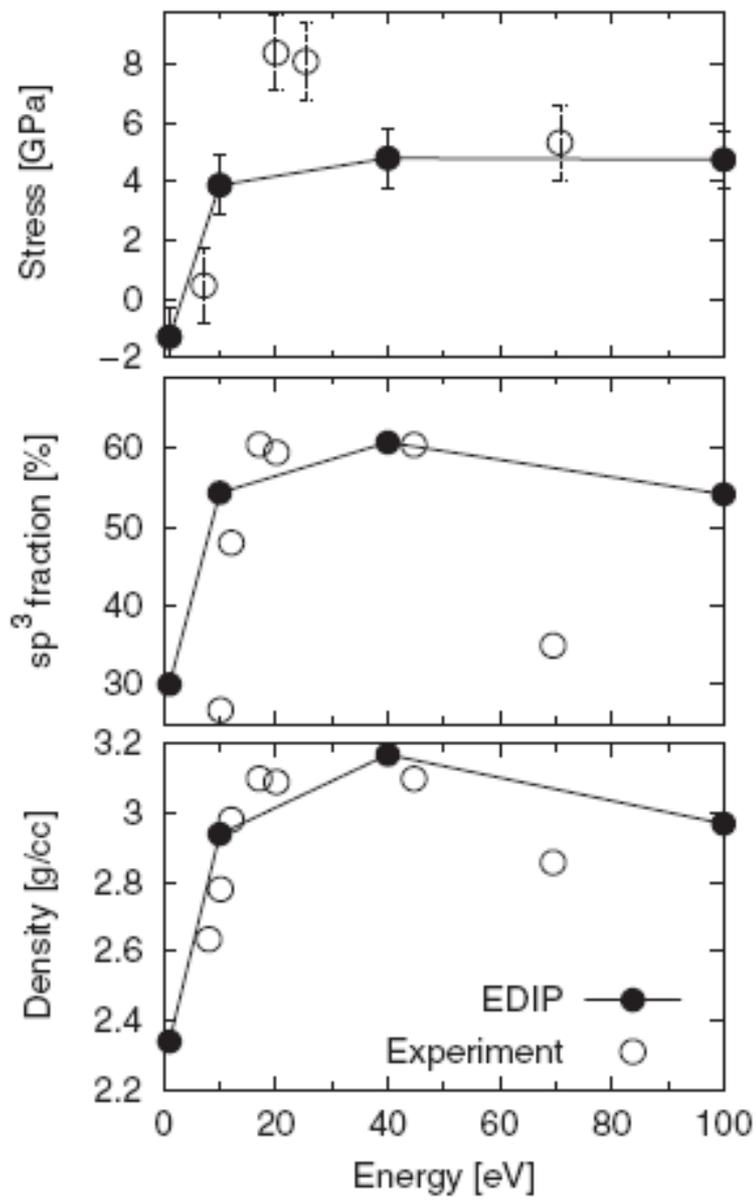

Fig. 5 Graves and Brault

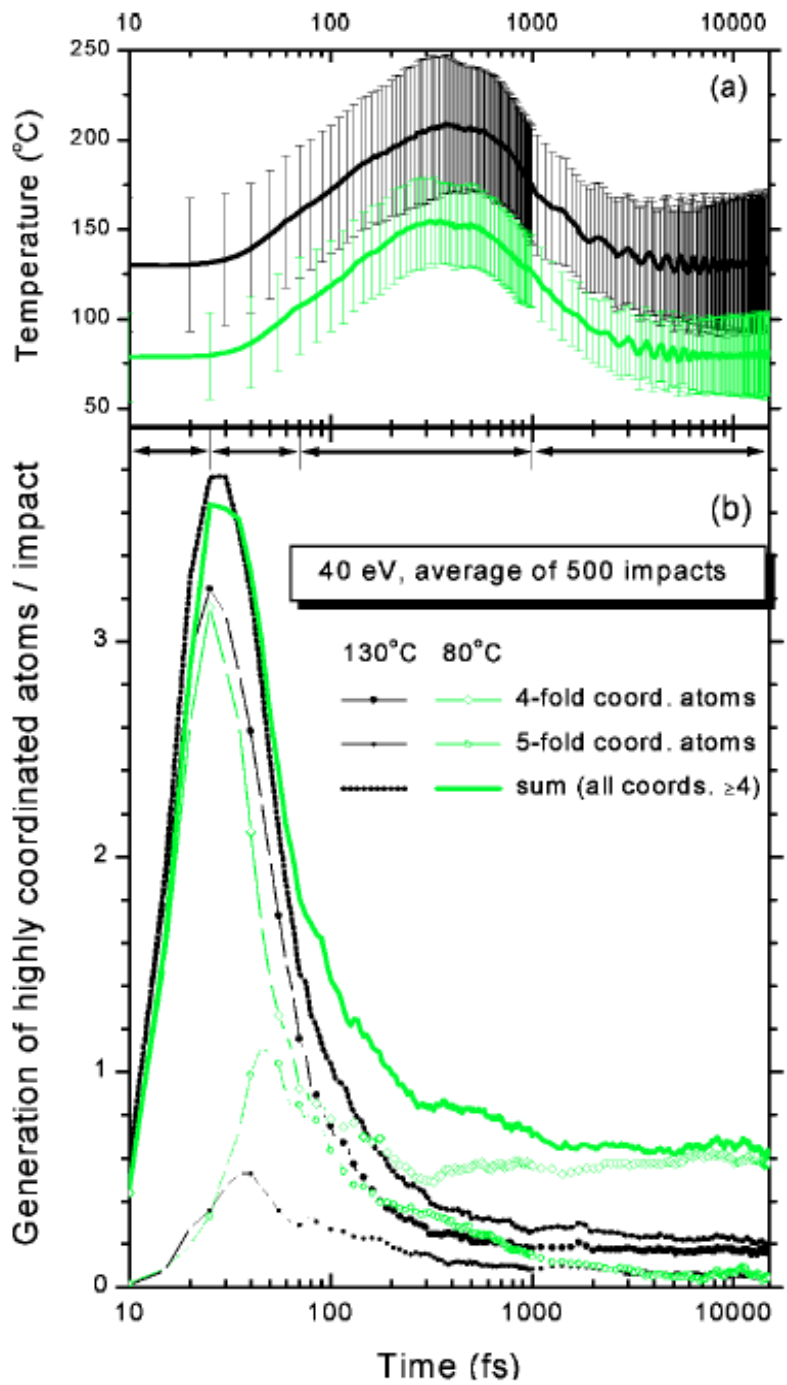

Fig. 6 Graves and Brault

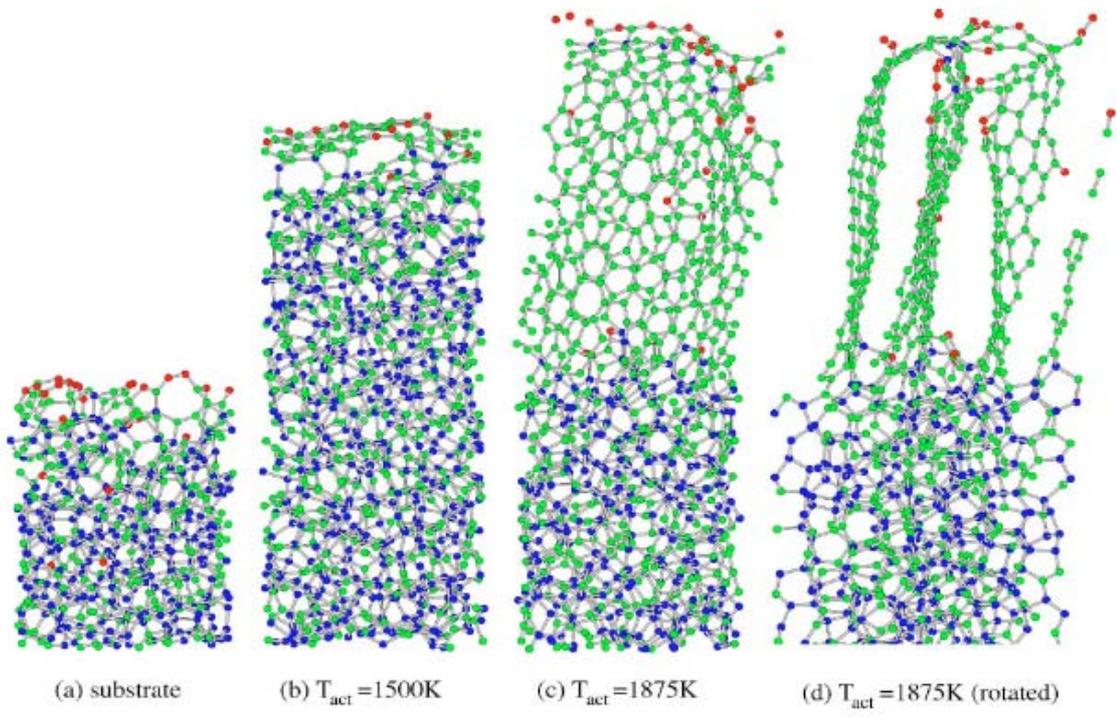

(a) substrate  (b) $T_{act} = 1500K$  (c) $T_{act} = 1875K$  (d) $T_{act} = 1875K$ (rotated)

Fig 7. Graves and Brault

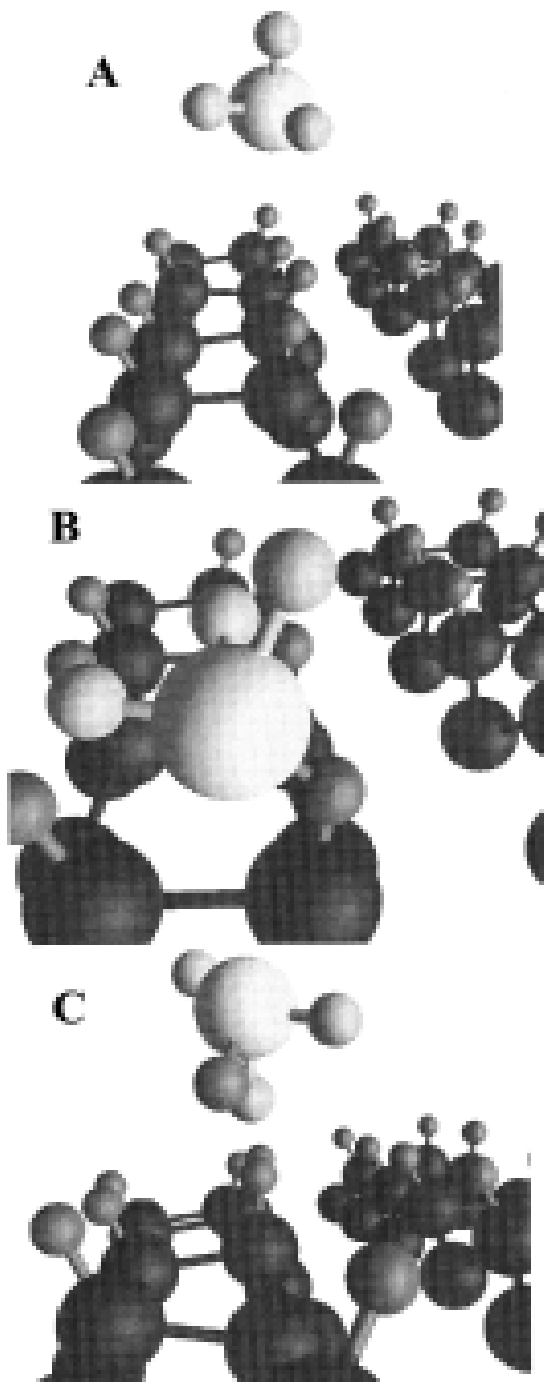

Fig. 8 Graves and Brault

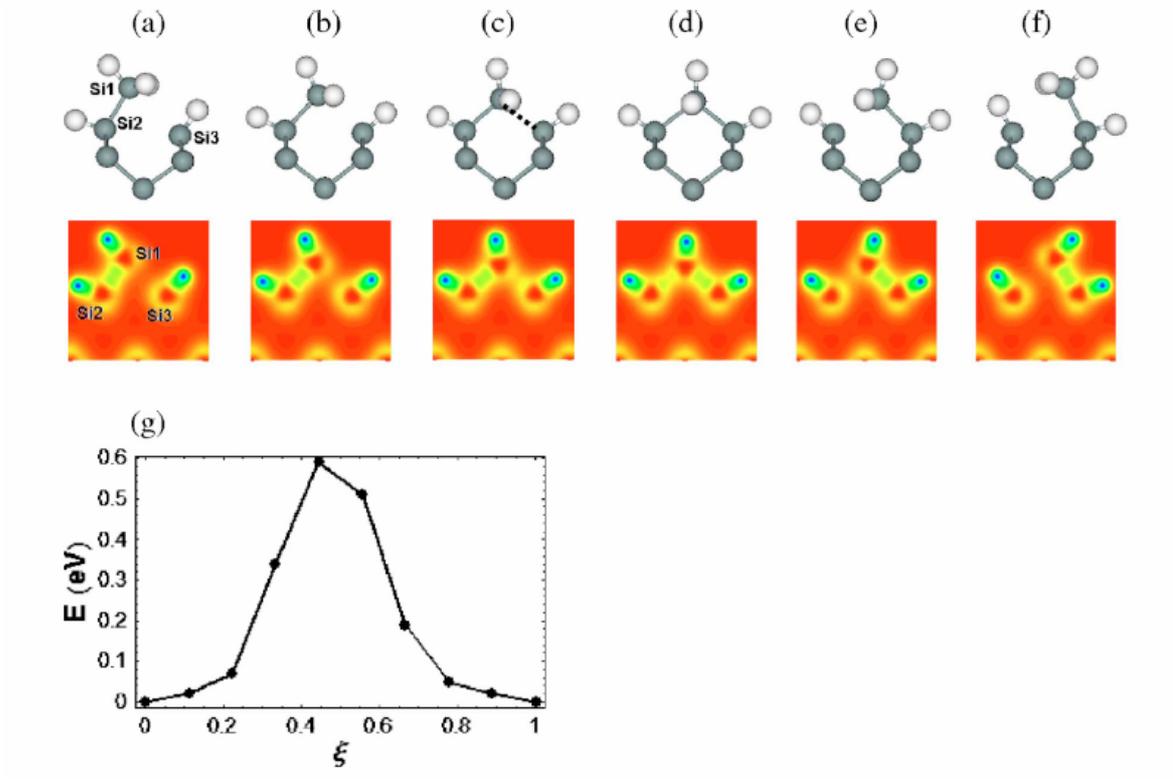

Fig. 9 Graves and Brault

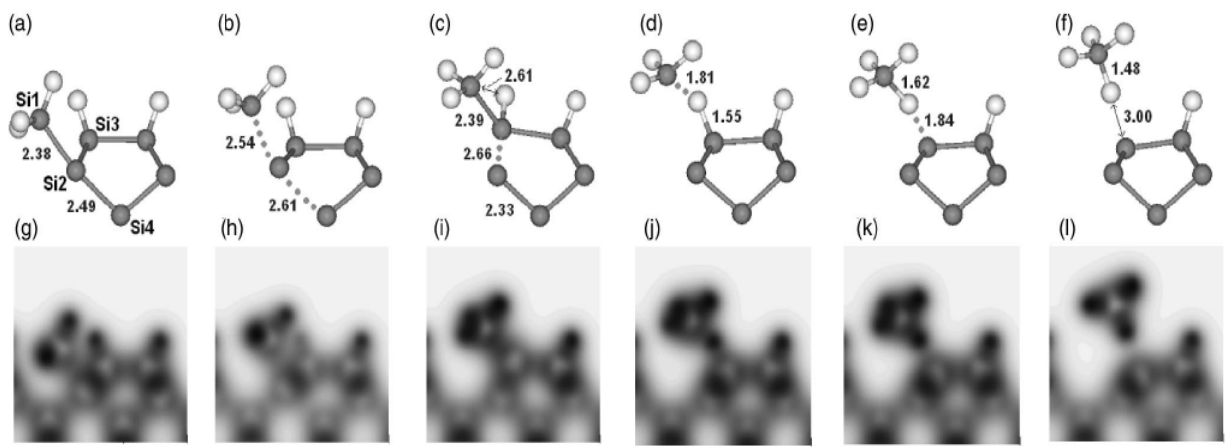

Fig. 10 Graves and Brault

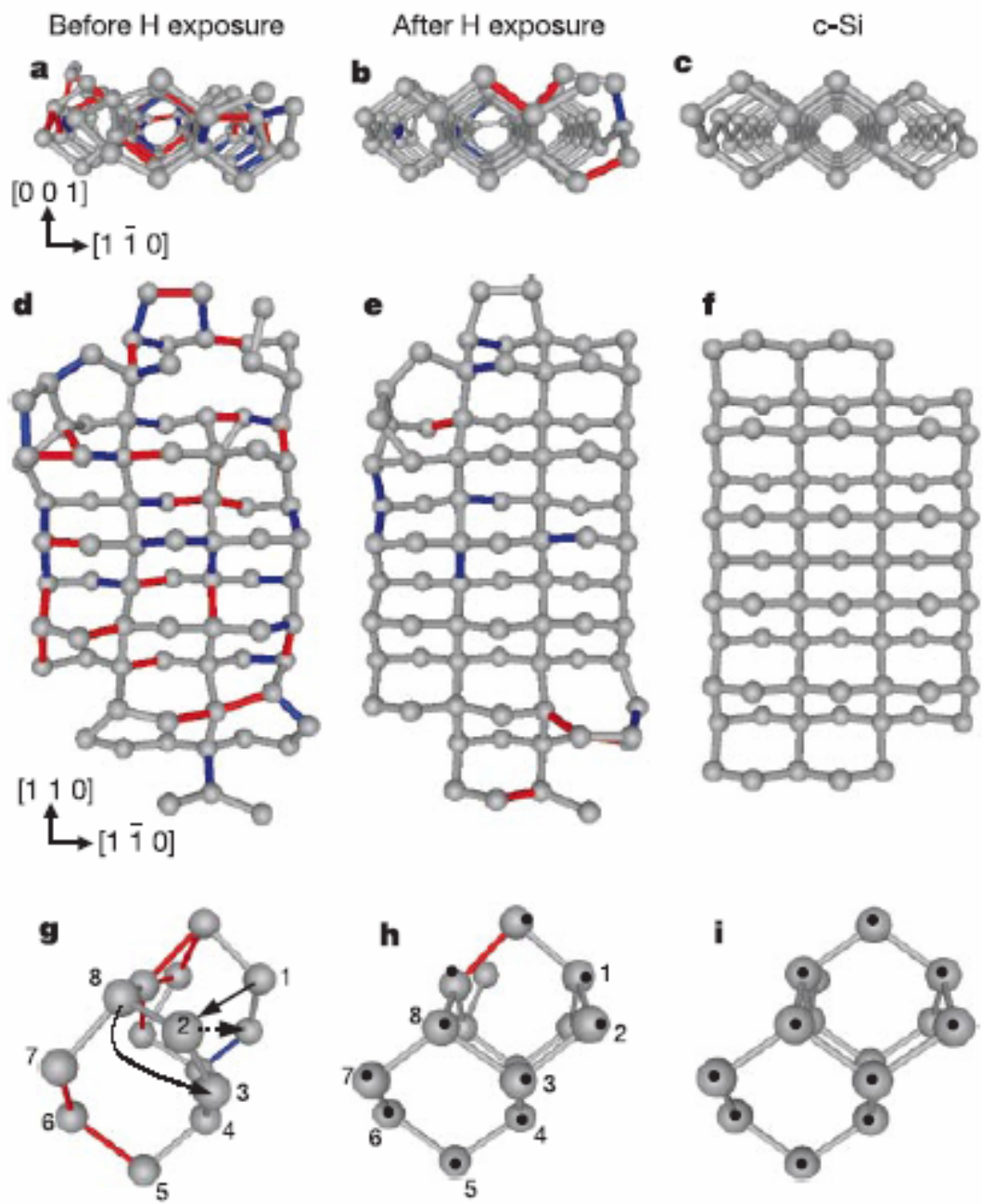

Fig. 11 Graves and Brault

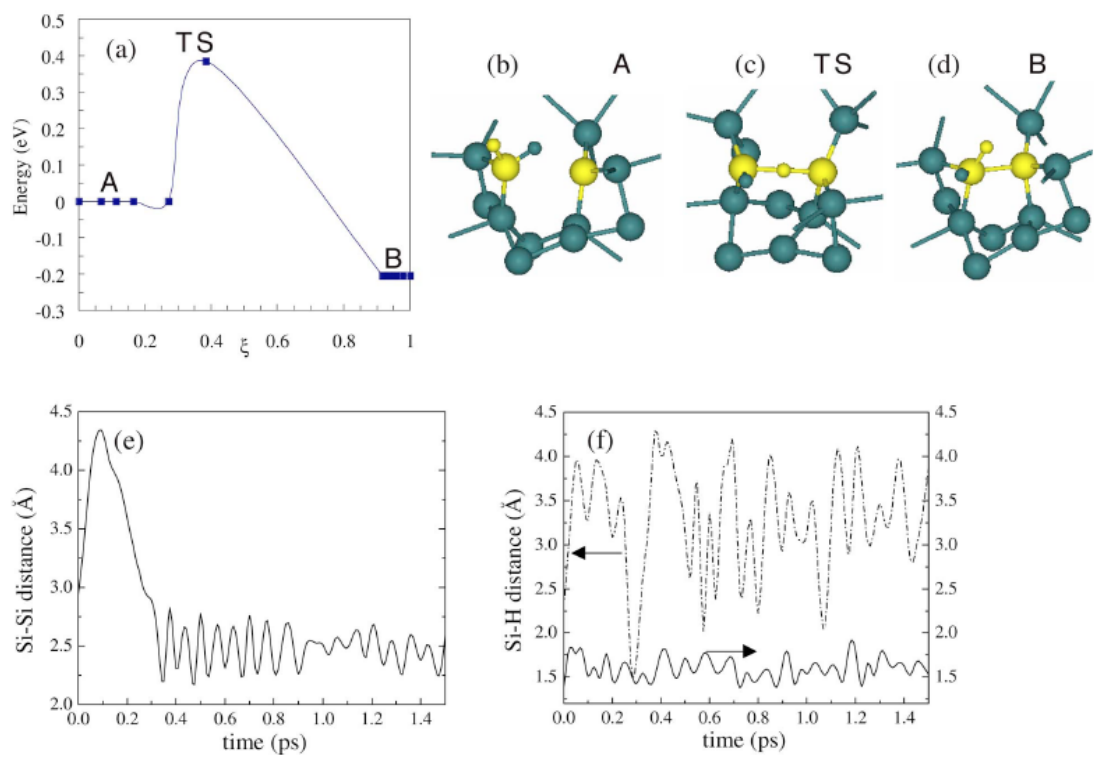

Fig. 12 Graves and Brault

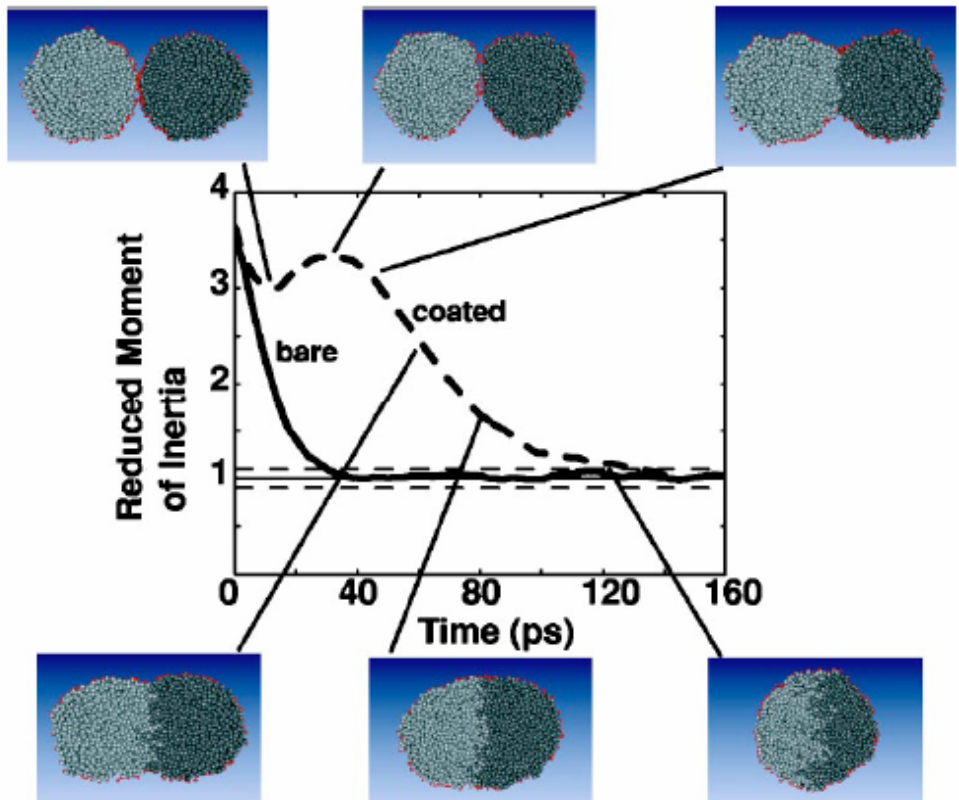

Fig. 13 Graves and Brault

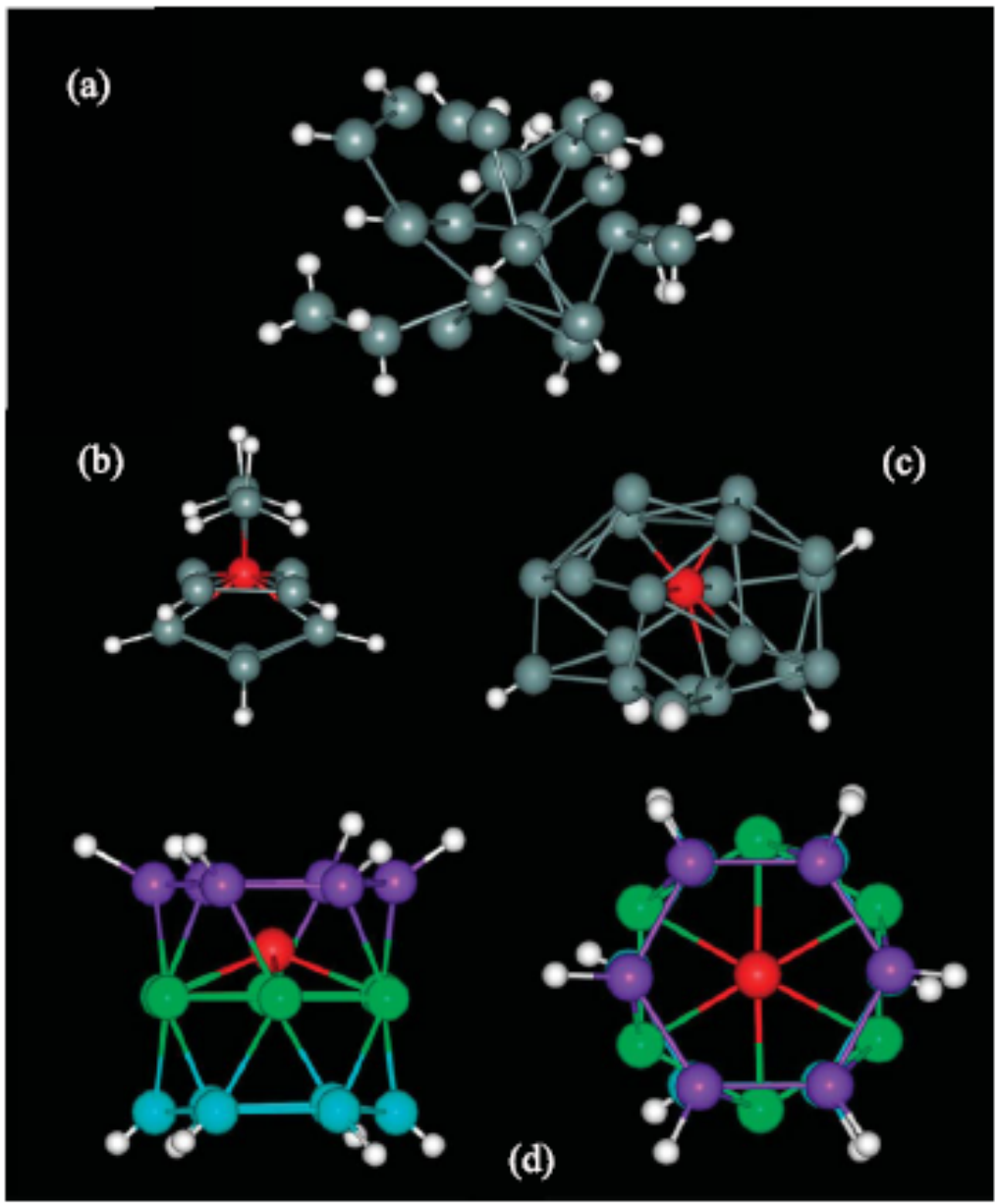

Fig. 14 Graves and Brault

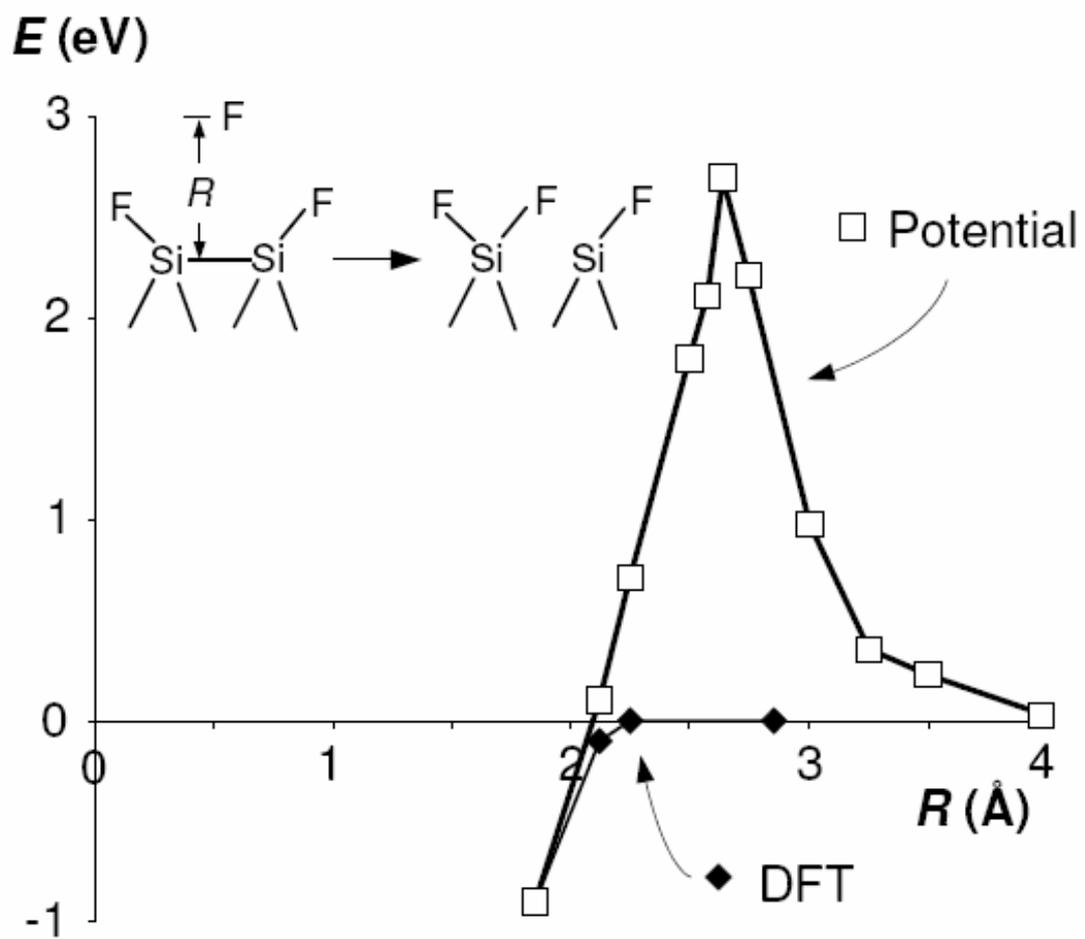

Fig. 15 Graves and Brault

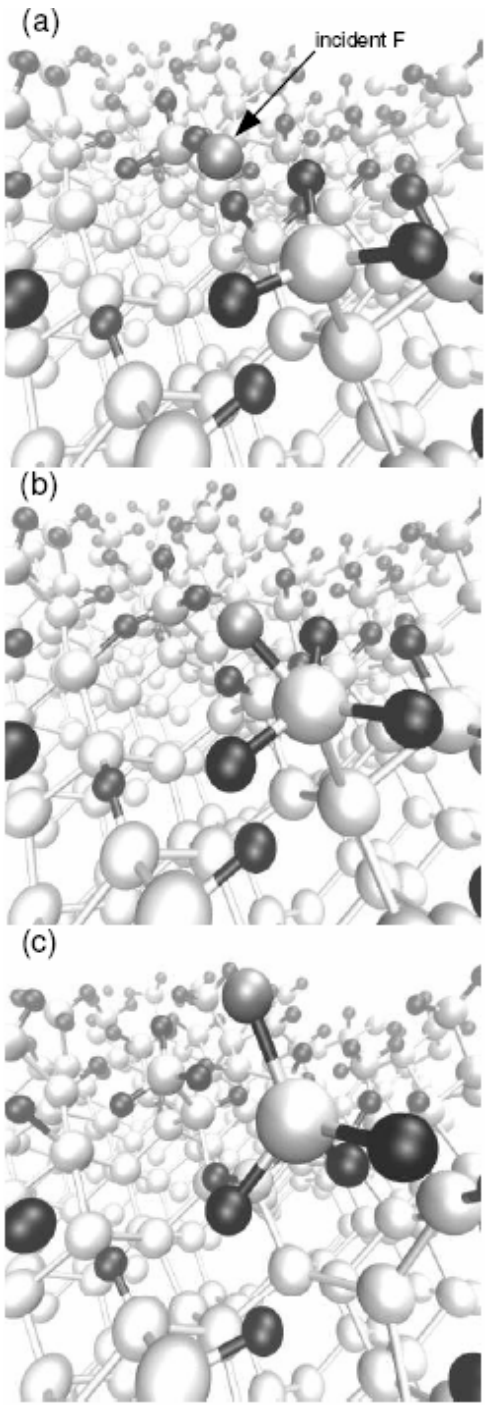

Fig. 16 Graves and Brault

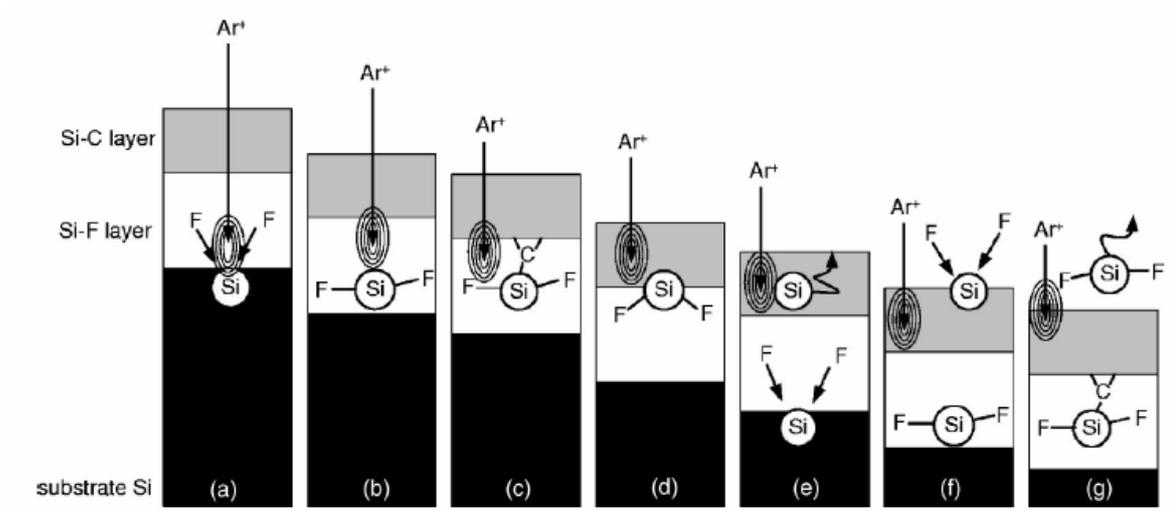

Fig. 17 Graves and Brault

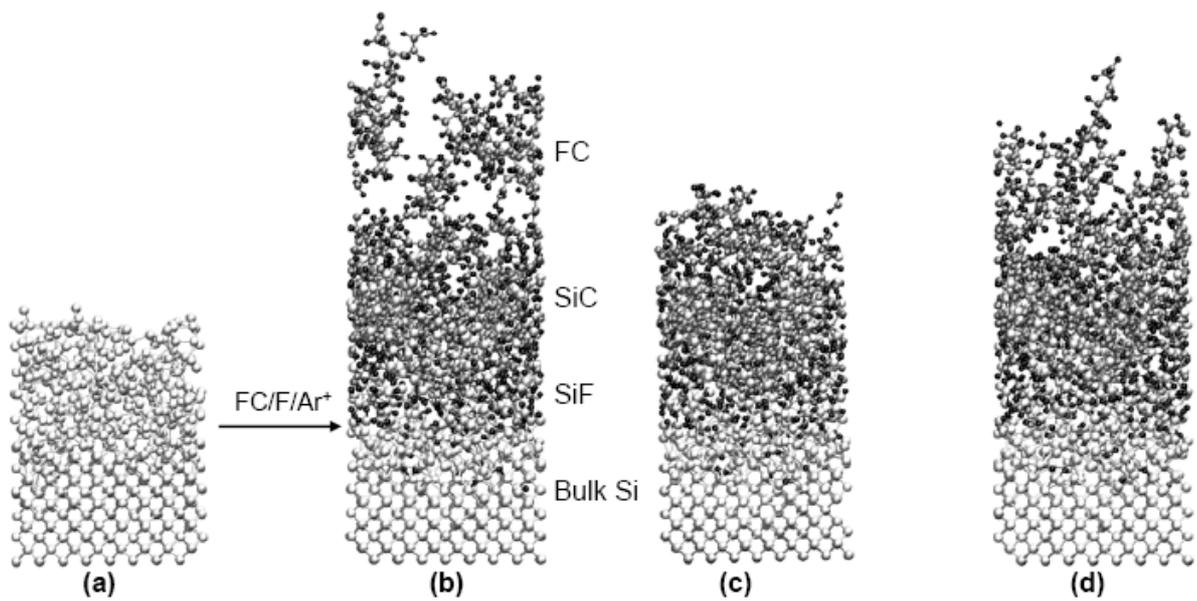

Fig. 18 Graves and Brault

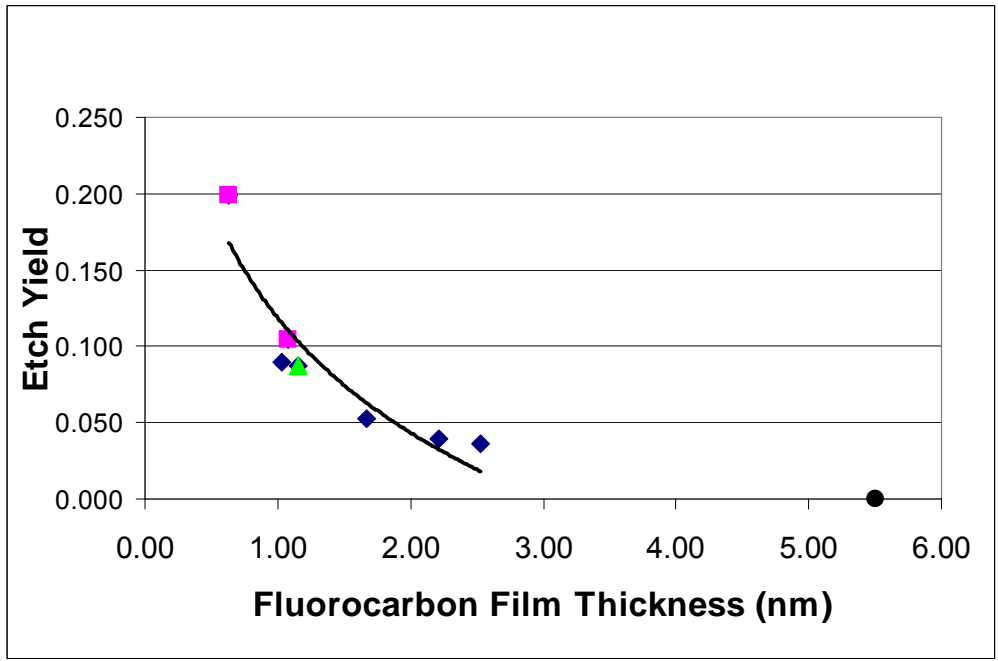

Fig. 19 Graves and Brault

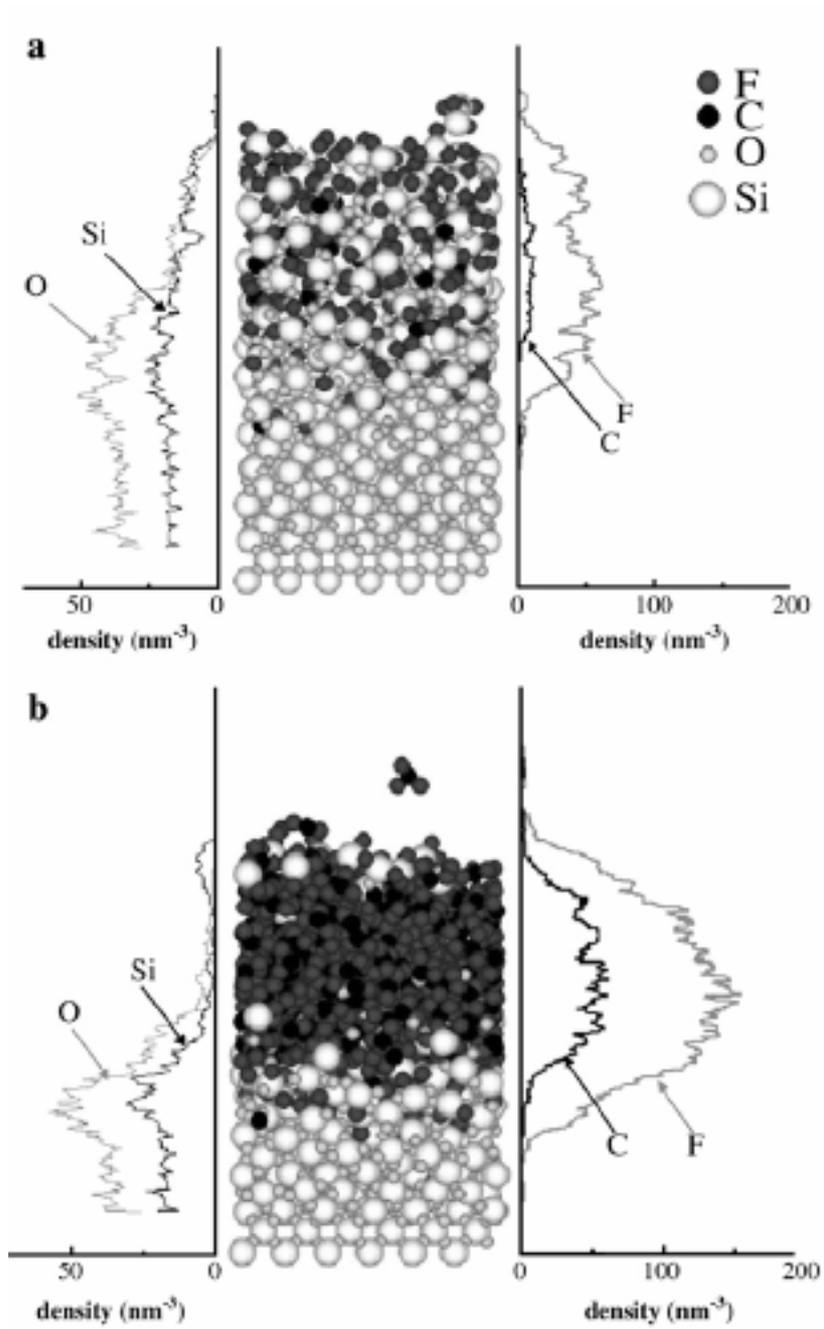

Fig. 20 Graves and Brault

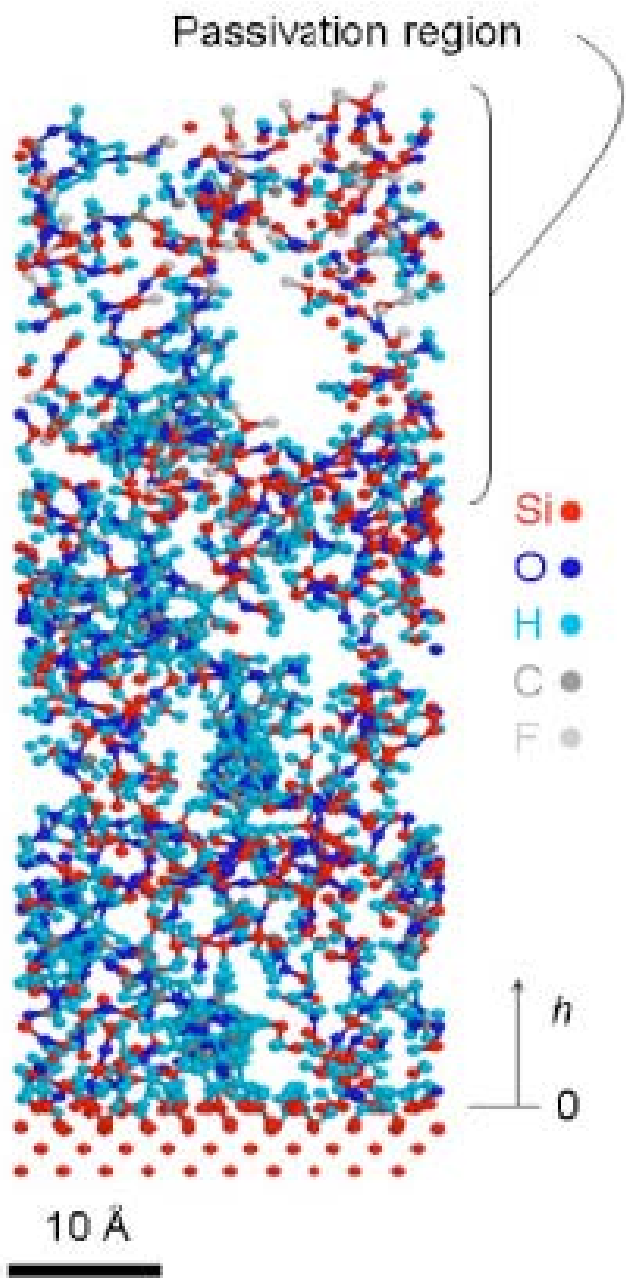

Fig. 21 Graves and Brault

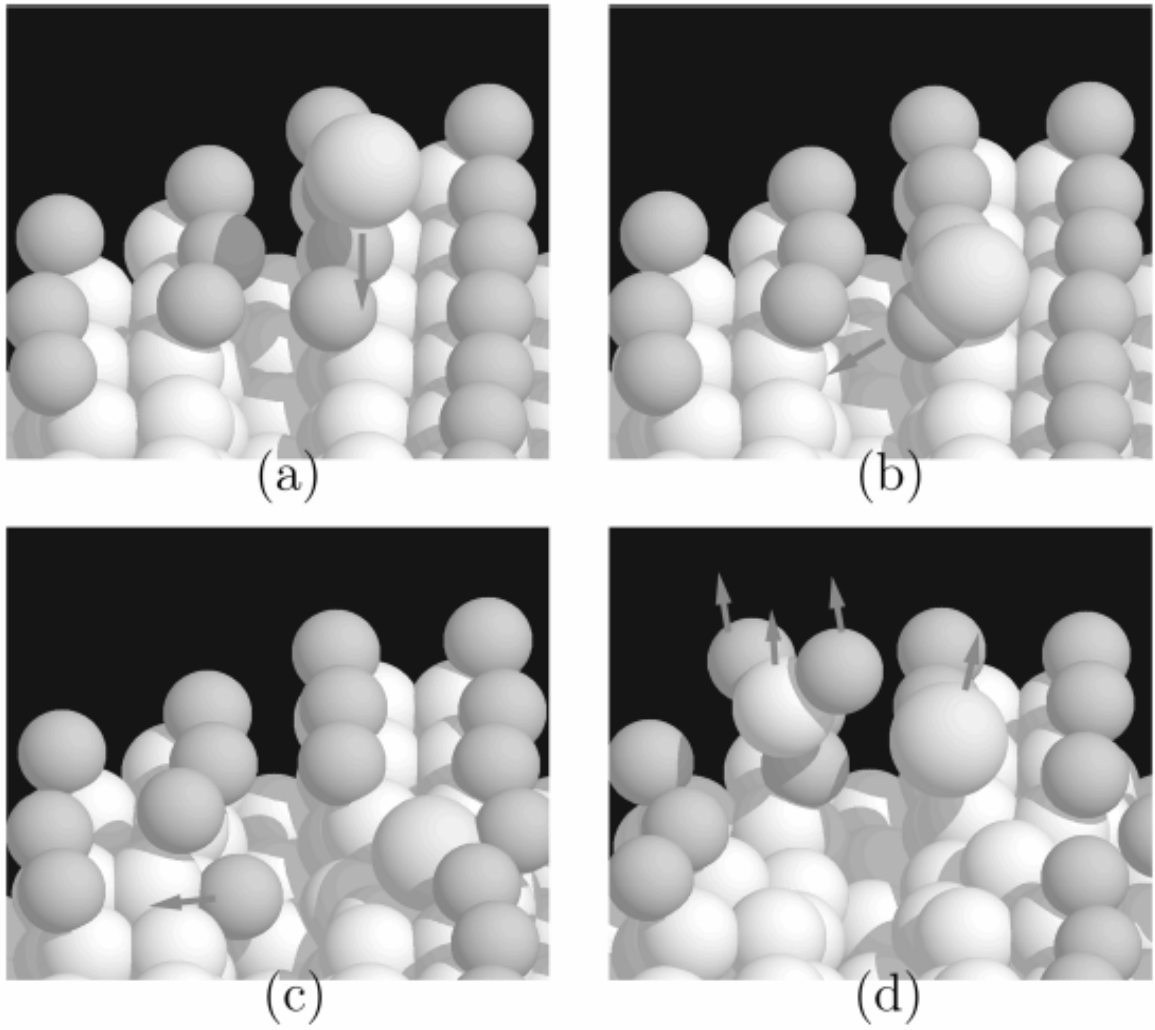

Fig. 22 Graves and Brault